%% file: main.tex
\documentclass[%
 reprint,
 amsmath,amssymb,
 aps,
]{revtex4-2}

\usepackage[utf8]{inputenc} 
\usepackage[T1]{fontenc}    
\usepackage{hyperref}       
\usepackage{url}            
\usepackage{booktabs}       
\usepackage{amsfonts,amsmath,amssymb}       
\usepackage{nicefrac}       
\usepackage{microtype}      
\hypersetup{colorlinks=true,linkcolor=blue,citecolor=red!60!blue}
\usepackage[cmyk]{xcolor}
\usepackage{graphicx}

\usepackage{caption}
\usepackage{subcaption}

\input{commands.tex}

\begin{document}

\title{Neuronal Temporal Filters as Normal Mode Extractors}


\author{Siavash Golkar}
\thanks{Equal contribution}
\affiliation{Center for Computational Neuroscience, Flatiron Institute}
\author{Jules Berman}
\thanks{Equal contribution}
\affiliation{Center for Computational Neuroscience, Flatiron Institute}
\author{David Lipshutz}
\affiliation{Center for Computational Neuroscience, Flatiron Institute}
\author{Robert Mihai Haret}
\affiliation{University Medical Center Göttingen, Department of Ophthalmology and Bernstein Center for Computational Neuroscience Göttingen}
\author{Tim Gollisch}
\affiliation{University Medical Center Göttingen, Department of Ophthalmology and Bernstein Center for Computational Neuroscience Göttingen}
\author{Dmitri B.\ Chklovskii}
\affiliation{Center for Computational Neuroscience, Flatiron Institute}
\affiliation{Neuroscience Institute, NYU Medical Center}
   
\date{\today}

\begin{abstract}
To generate actions in the face of physiological delays, the brain must predict the future. Here we explore how prediction may lie at the core of brain function by considering a neuron predicting the future of a scalar time series input. Assuming that the dynamics of the lag vector (a vector composed of several consecutive elements of the time series) are locally linear, Normal Mode Decomposition  decomposes the dynamics into independently evolving (eigen-)modes allowing for straightforward prediction. We propose that a neuron learns the top mode and projects its input onto the associated subspace. Under this interpretation, the temporal filter of a neuron corresponds to the left eigenvector of a generalized eigenvalue problem. We mathematically analyze the operation of such an algorithm on noisy observations of synthetic data generated by a linear system. Interestingly, the shape of the temporal filter varies with the signal-to-noise ratio (SNR): a noisy input yields a monophasic filter and a growing SNR leads to multiphasic filters with progressively greater number of phases. Such variation in the temporal filter with input SNR resembles that observed experimentally in biological neurons.
\end{abstract}

\maketitle


\section{Introduction}

The brain must generate behavior in the face of physiological delays on multiple levels: from sensory transduction to axonal conduction to synaptic transmission and muscle activation. To compensate for such delays, it would be useful even for individual neurons to predict future inputs. Not surprisingly, the paradigm of optimal prediction has been used to derive normative models of neurons. Such models can be loosely classified into two categories \citep{singer2018sensory,chalk2018toward}: predictive coding \citep{srinivasan1982predictive,rao1999predictive,huang2011predictive, gregor2012lattice} and predictive information \citep{tishby2000information,bialek2006efficient}. 

\begin{figure}[h]
     \centering
    \includegraphics[width=0.42\textwidth]{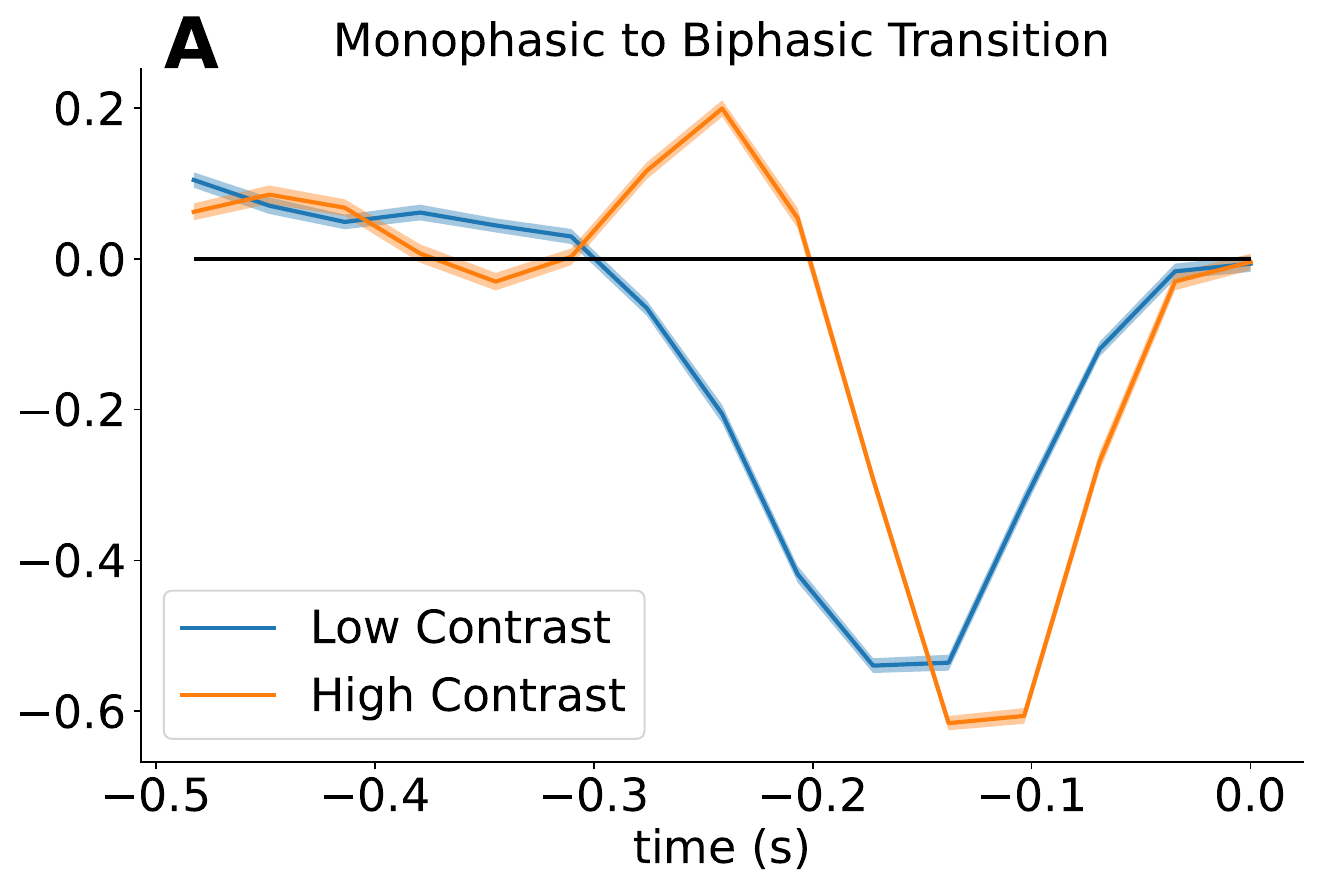}
    \includegraphics[width=0.42\textwidth]{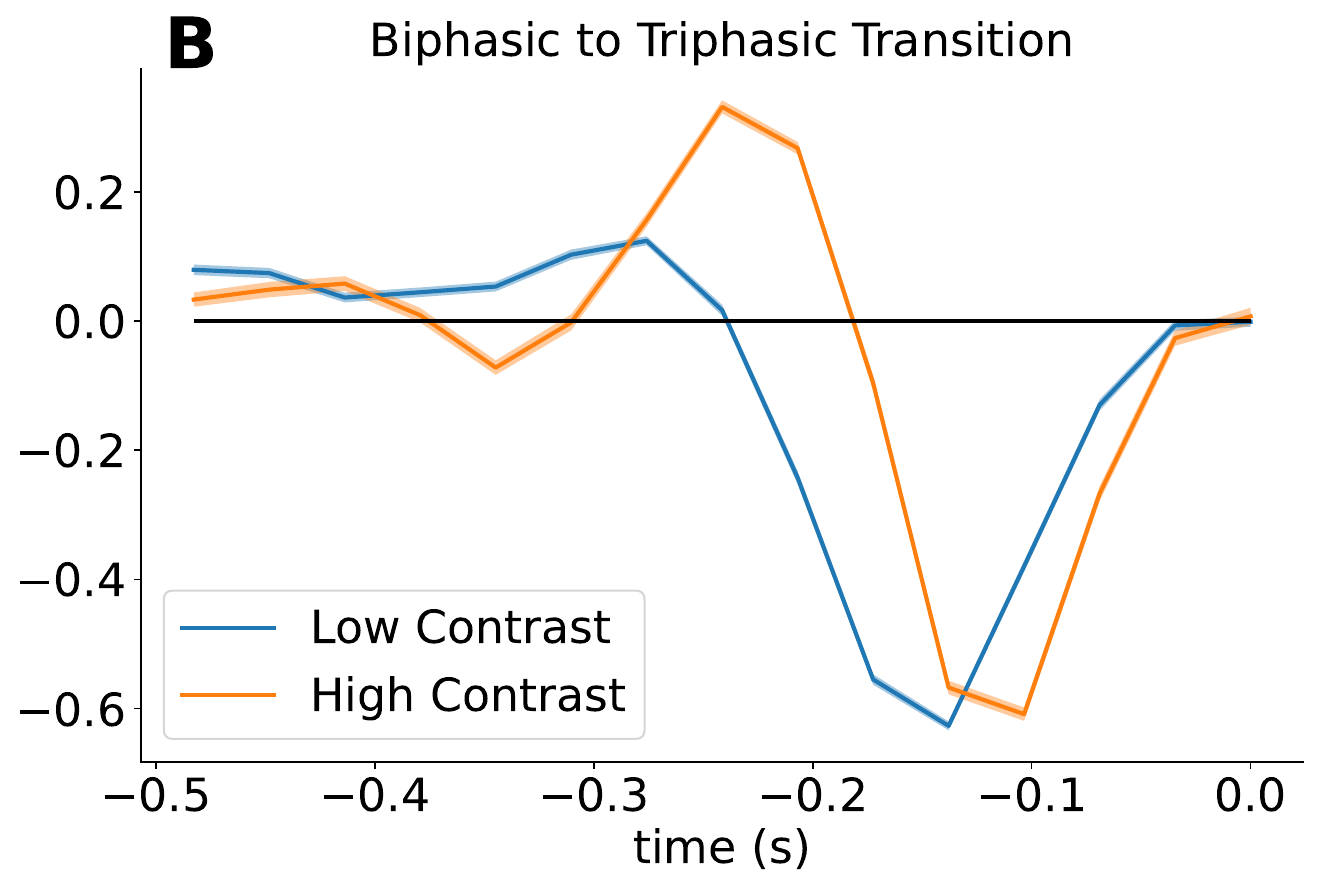}
    \includegraphics[width=0.42\textwidth]{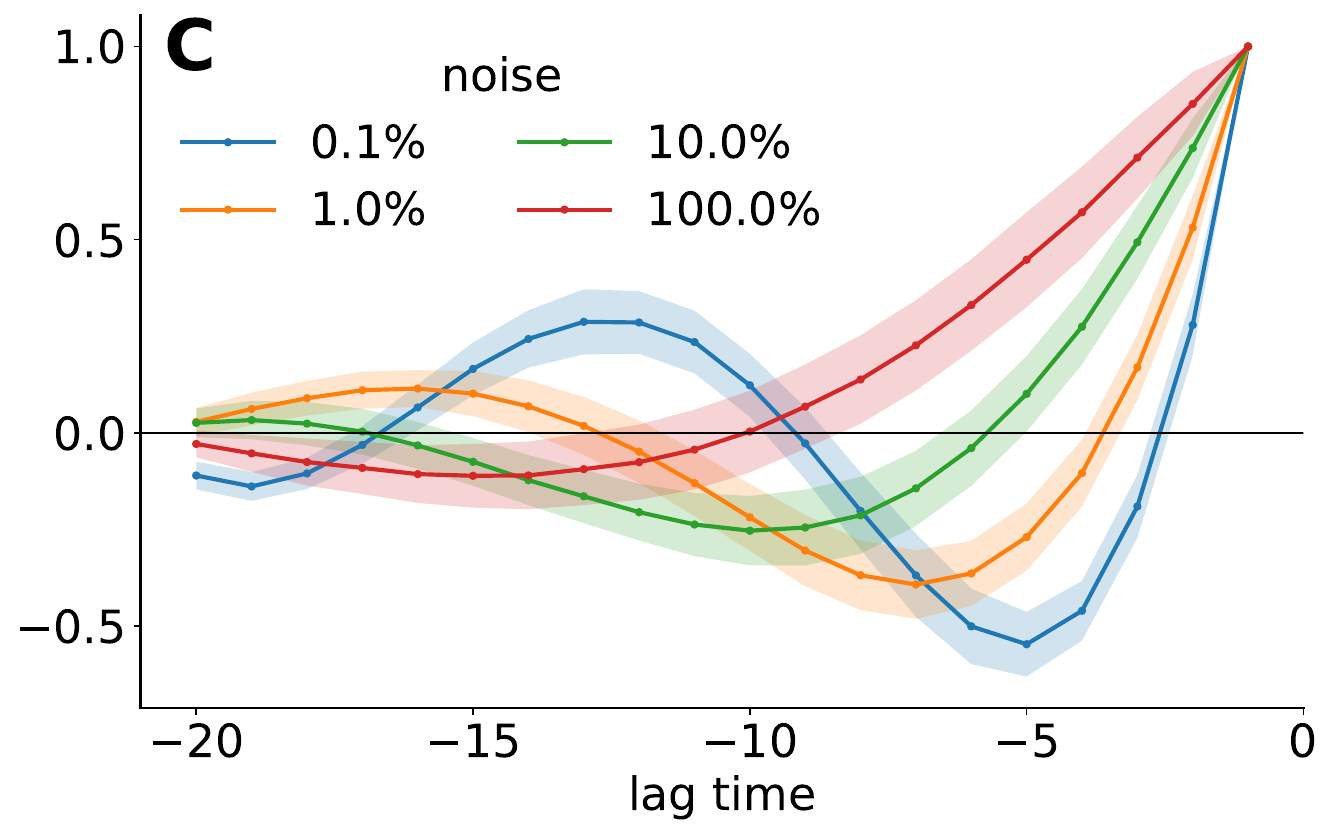}
    \label{fig:samples}
        \caption{Neuronal temporal filters. (A,B) Experimentally obtained by Spike Triggered Average (STA) from two retinal ganglion cells stimulated by white noise of different contrast. (C) Top left eigenvectors obtained in our framework for different levels of noise in a synthetic example. Shading shows standard deviation.  Whereas theoretical filters peak when the neuron spikes, experimental ones peak before that. This artifact is due to signal processing delays from photo-absorption to the ganglion cell firing and the smoothness of experimental filters resulting from the low-pass nature of biophysical processes.}
    \label{fig:intro}
\end{figure}

In predictive coding, a neuron computes an optimal prediction of the signal and subtracts it from the actual value, transmitting prediction error downstream \citep{srinivasan1982predictive,rao1999predictive,huang2011predictive}. This results in an optimal linear filter whose shape depends on the statistics of the input. Predictive coding approximately accounts for the change in the temporal receptive field with the input SNR (Fig.~\ref{fig:intro}A) but fails to reproduce the exact shape of the filter because it has a narrow unitary peak at the exact time of the actual signal (usually present time) \citep{srinivasan1982predictive}. Such peak can be smoothed by adding a de-noising objective but that introduces additional parameters to the model. 

In predictive information, a neuron filters the past signal to preserve only the part which is most informative about the future \citep{tishby2000information,bialek2006efficient,chalk2018toward}. Such framework expands the assortment of temporal filters depending on the various parameters such as the balance between the past/output and output/future mutual information. Recently, it was shown that multi-phasic filter can be derived using this formalism as well~\cite{retinal-mosaics}.

A shortcoming of both approaches is that they require setting the time point relative to the present for which the neuron generates prediction. Such formulation of the prediction problem does not seem suitable when neurons might have different amounts of delay, or the goal is to predict (and act upon) a trend or a mode of the signal that spans multiple time points in the future. 
Moreover, both approaches view the neuronal filter as a trade off between different mathematical terms in the optimization objective and therefore depend on the relative weighting of these terms. Therefore, the shape of the temporal kernel depends on these - as well as other - hyperparameters of the model.

%
An alternative approach to prediction assumes that the sensory stimuli are generated by a (generally nonlinear) dynamical system. If such dynamics can be learned by the brain from previously seen trajectories, then prediction can be cast as identifying long-living (growing or slow) modes (or manifolds) in the input and extrapolating them into the future. To take the first step in this direction, we take advantage of the fact that in the vicinity of a hyperbolic fixed point the invariant manifolds can be well approximated by the invariant subspaces of the linearized dynamics (see e.g. Sec.~19.12a of~\cite{wiggins2006introduction}). Therefore, a neuron that identifies the least decaying linear mode can identify an invariant manifold of a fixed point and generate a non-trivial prediction.

Here, we explore the hypothesis that a neuron extracts the least decaying mode from the incoming scalar time series as in Normal Mode Decomposition \citep{Landau1976Mechanics}. To cast this problem in the language of dynamics, we construct time-lag vectors by windowing the scalar time series. We assume that the system is in the vicinity of a hyperbolic fixed point and identify the linear dynamics of the lag vectors. The eigen-decomposition of the linear transition matrix yields the eigenmodes that evolve independently as the right eigenvectors and the corresponding linear filters as the left eigenvectors. Therefore, the top left eigenvector yields a linear filter that projects onto the most dominant mode, representing neuronal output. Thus, unlike the previous predictive coding and predictive information approaches which optimize prediction for a certain time in the future, in our proposal, a neuron learns a rank-1 approximation of the dynamical system generating the input and outputs the top mode which develops into the future in a predictable way.

Our theoretical proposal makes a prediction that can be compared with experimental data. We find that the shape of the top left eigenvector depends on the input SNR: monophasic filter at low SNR with adding phases for growing SNR (Fig.~\ref{fig:intro}C). This prediction is reminiscent of the SNR-induced variation in the form of the Spike-Triggered Average (STA) of various types of biological neurons (Fig.~\ref{fig:intro}AB)\cite{srinivasan1982predictive,baccus2002fast,liu2015spike,chander2001adaptation,kim2001temporal,mainen1995reliability}. Although this does not prove our proposal it suggests that this approach may be a step in the right direction.

\section{Normal Mode Decomposition (NMD)}\label{sec:DMD}
In this Section, we review NMD \citep{Landau1976Mechanics} in the context of a neuron with a Single Input and a Single Output (SISO)~\footnote{In a neuron receiving synapses from multiple sources this would correspond to considering the total synaptic current as a scalar input.}. 

\subsection{Problem statement and linearization}

Let $\{x(t)\}_{t=1,2,\dots}$ be a scalar time series satisfying the non-linear relation
\begin{equation}\label{eq:nl_dyn}
    x(t+1)=f(x(t),\dots,x(t-n)),\qquad t\ge n+1,
\end{equation}
where $f:\R^n\to\R$ is a continuously differentiable function and $n\ge1$ is referred to as the \textit{lag}. We say $x^\ast\in\R$ is a \textit{fixed point} if $x^\ast=f(x^\ast,\dots,x^\ast)$. We embed a scalar time series $x(t)$, $t=1,\dots,T+n$, into a sequence of lag vectors $\x_t=[x(t-n+1),\dots,x(t)]^\top\in\R^n$. The dynamics of the first $n-1$ components of the lag vector is given by a simple shift in the time-step (i.e. $x(t-i)\to x(t-i+1)$) and the dynamics of the final component $x(t)$ is given by Eq.~\eqref{eq:nl_dyn} (i.e. $x(t)\to f(x(t),\dots,x(t-n))$). These can be summarized as $\x_{t+1}=\F(\x_t)$, where the function $\F$ is given by
\begin{equation}
\label{dyn_nl}
    \F(\x_t) = \F(\begin{bmatrix}
        x(t-n+1) \\ \vdots \\ x(t-1) \\ x(t)
    \end{bmatrix})\equiv\begin{bmatrix}
        x(t-n+2) \\ \vdots \\ x(t) \\ \\ f(x(t),\dots,x(t-n))
    \end{bmatrix}
\end{equation}
We assume that the dynamics are dominated by the presence of a hyperbolic fixed point at $\x^*$. 
We approximate dynamics \eqref{dyn_nl} by expanding around this fixed point. Defining $\delta \x_t = \x_t - \x^*$, we have:
\begin{align*}
    \x_{t+1}&=\F(\x^*)+ \nabla\F(\x^*)\delta\x_t+o(\|\delta\x_t\|)\\
    &\approx \x^* + \nabla\F(\x^*)\delta\x_t,
\end{align*}
where we have used the fact that $\F(\x^*)=\x^*$ and $\nabla\F(\x^*)$ is the Jacobian evaluated at $\x^*$. After subtracting $\x^*$ from both sides:
\begin{align*}
    \delta \x_{t+1}\approx\nabla\F(\x^*)\delta\x_t
\end{align*}
Letting $\A:=\nabla\F(\x^*)$ and assuming without loss of generality that $\x^*=
\bf0$, we have the following linear approximation to the nonlinear dynamics in equation \eqref{dyn_nl} near the fixed point:
\begin{equation}
\label{dyn}
    \x_{t+1}=\A\x_t.
\end{equation}
From the definition of the function $\F$ in Eq.~\eqref{dyn_nl}, we see that $\A=\nabla\F(\bf 0)$ takes a companion matrix form~\citep{bellman1997introduction}
\begin{equation}\label{eq:A_def}
    \A=
    \begin{bmatrix}
        0 & 1 &  &   \\
         & 0 & 1 &   \\
         &  & \ddots & \ddots &   \\
         &  &  &  0 & 1 \\
        c_1 & c_2 & \hdots & &c_n
    \end{bmatrix}.
\end{equation}
where $c_i=\partial f(\x^\ast)/\partial x^i$.

\subsection{Eigendecomposition of the dynamics}
The matrix $\A\in \R^{n\times n}$ is generically a non-normal matrix possessing an eigendecomposition:
\begin{equation}
\label{DMD}
    \A=\sum_i \lambda_i\w_i \v_i^\top \;\;\; {\rm s.t.} \;\;\; \v_i^\top\w_j=\delta_{ij},
\end{equation}
where $\lambda_i$ are the eigenvalues and $\w_i$ (resp. $\v_i^\top$) are the right (resp. left) eigenvectors.
Here, we consider
 matrix $\mathbf{A}$ describing a hyperbolic fixed point ($\lambda_i\neq 1$). Moreover, we assume, that all eigenvalues are real and are sorted in descending order \mbox{$\lambda_i\geq \lambda_{i+1}$}. We also assume that the top mode is non-degenerate (i.e. $\lambda_1>\lambda_2$).


The normal modes are found by eigendecomposition of~$\A$.  As a matrix of companion form, $\A$ is diagonalized using the Vandermonde matrix and its inverse~\citep{bellman1997introduction}:
\begin{align*}
    \A=\V\Lam\V^{-1},
\end{align*}
where $\V$ is the Vandermonde matrix
\begin{align*}
    \V:=\begin{bmatrix}
        1&1&\cdots&1\\
        \lambda_1&\lambda_2&\cdots & \lambda_n\\
        \vdots& \vdots & & \vdots\\
        \lambda_1^{n-1} & \lambda_2^{n-1} & \cdots & \lambda_n^{n-1}
    \end{bmatrix}
\end{align*}
and $\Lam:=\text{diag}(\lambda_1,\dots,\lambda_n)$ is the diagonal matrix of eigenvalues $\lambda_i$, which are the roots of the characteristic polynomial with coefficients $c_i$. The left eigenvectors of $\A$, are given by the rows of the inverse of the Vandermonde matrix $\V^{-1}$. It is easily verified that the right eigenvectors of $\A$, corresponding to the columns of the Vandermonde matrix, represent the individual modes of the dynamical system.  

\subsection{Projecting onto the dominant mode}
We are interested in finding the dominant exponential mode of the dynamics. As stated above, the dominant exponential mode is represented as the top right eigenvector of $\A$. Therefore, by projecting onto the top right eigenvector, we can isolate the dominant mode. 

Because of the bi-orthogonality of the left and the right eigenvectors, the top left eigenvector (henceforth referred to as $\v_1$) is orthogonal to all but the top right eigenvector. This allows us to use $\v_1$ as a projector that zeros out all but the most dominant mode of the dynamics. We therefore propose that it is the neuron's goal to learn the top left eigenvector and project its input onto this vector.

In order to learn the top left eigenvector, one approach would be to first find matrix $\A$ and then perform the eigen-decomposition on the inferred $\A$ matrix. This matrix can be found from data $x(t)$ by minimizing the mean squared error:
\begin{equation}
\label{mse}
    \min_{\A}\sum_t\left\lVert\x_{t+1}-\A\x_t\right\rVert^2.
\end{equation}
Eq.~\eqref{mse} can be solved via the following system of equations:
\begin{equation}
\label{mse-sol}
  \X_+\X^\top = \A\X\X^\top,
\end{equation}
where we introduce a matrix notation $\X=[\x_n,...,\x_{T+n-1}]$ and $\X_+=[\x_{n+1},...,\x_{T+n}]$.

In this paper, however, instead of directly solving for $\A$ via Eqs.~\eqref{mse} and \eqref{mse-sol}, we substitute the eigendecomposition, Eq.\eqref{DMD} and multiply both sides by $\v_i^\top$ on the left, leading to the equivalent generalized eigenvalue problem:
\begin{equation}\label{eq:gen_eig_prob}
  \v_i^\top \X_+\X^\top = \lambda_i\v_i^\top\X\X^\top.
\end{equation}
This allows us to circumvent the intermediate step of computing the $\A$ matrix explicitly.

\section{Detecting the dominant mode}
In this Section, we apply our formalism to time-series synthetically generated by a linear dynamical system of order $k$. 

\subsection{Problem formulation} 
\begin{figure*}[ht]
    \centering
    \includegraphics[width=0.24\textwidth, trim=5 0 5 0]{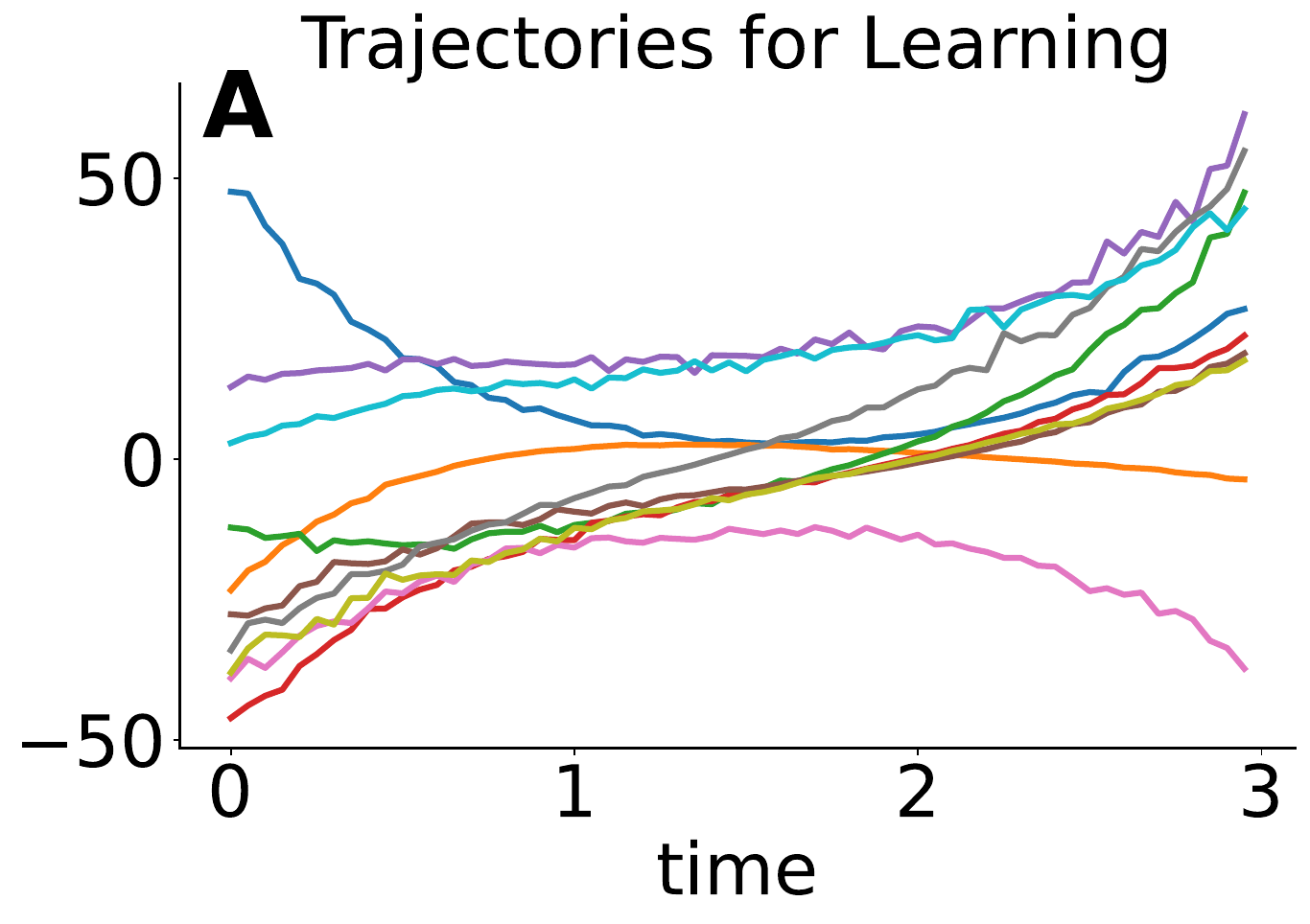}
    \hfill
    \includegraphics[width=0.24\textwidth, trim=5 0 5 0]{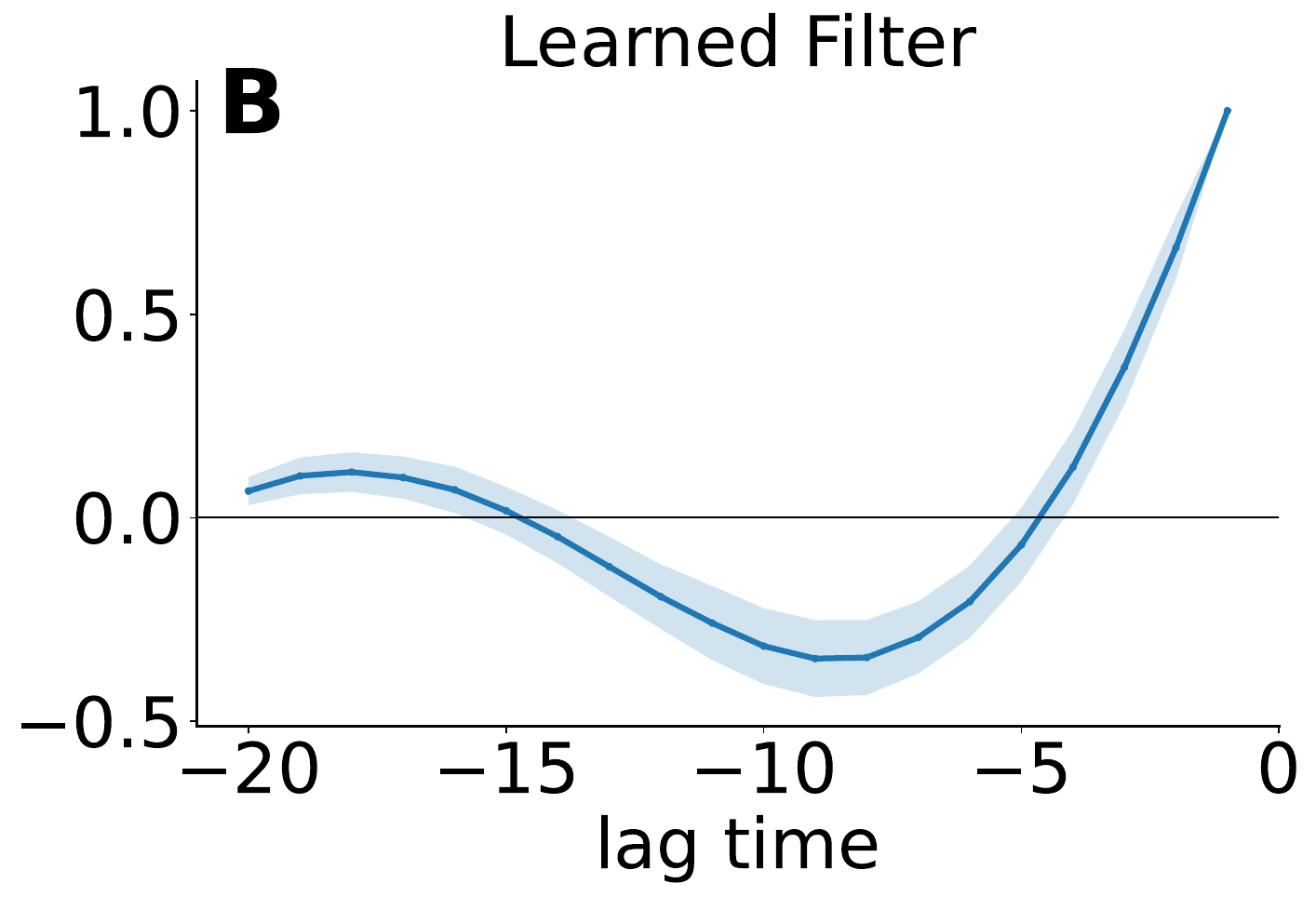}
    \hfill
    \includegraphics[width=0.24\textwidth, trim=5 0 5 0]{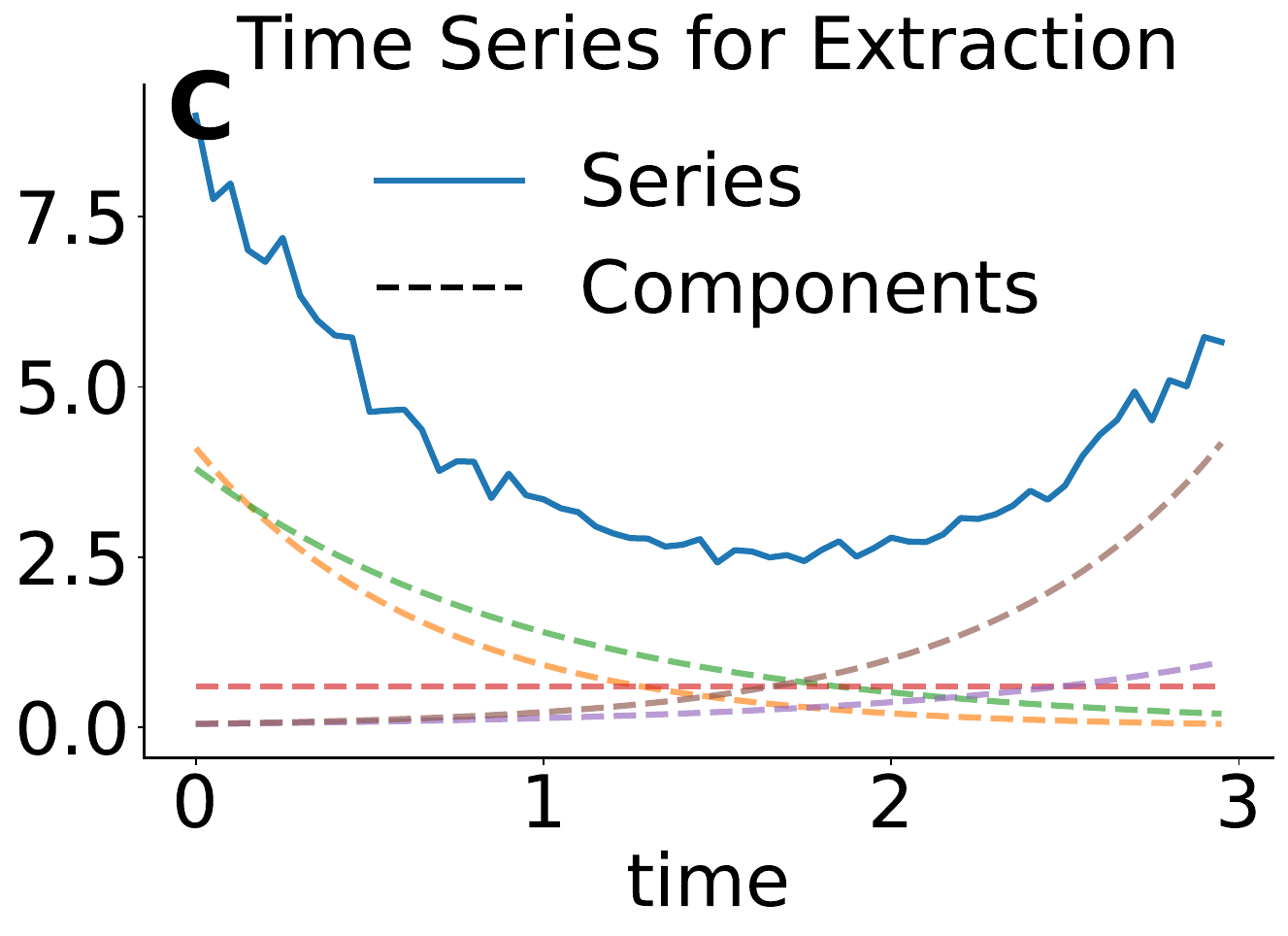}
    \hfill
    \includegraphics[width=0.24\textwidth, trim=5 0 5 0]{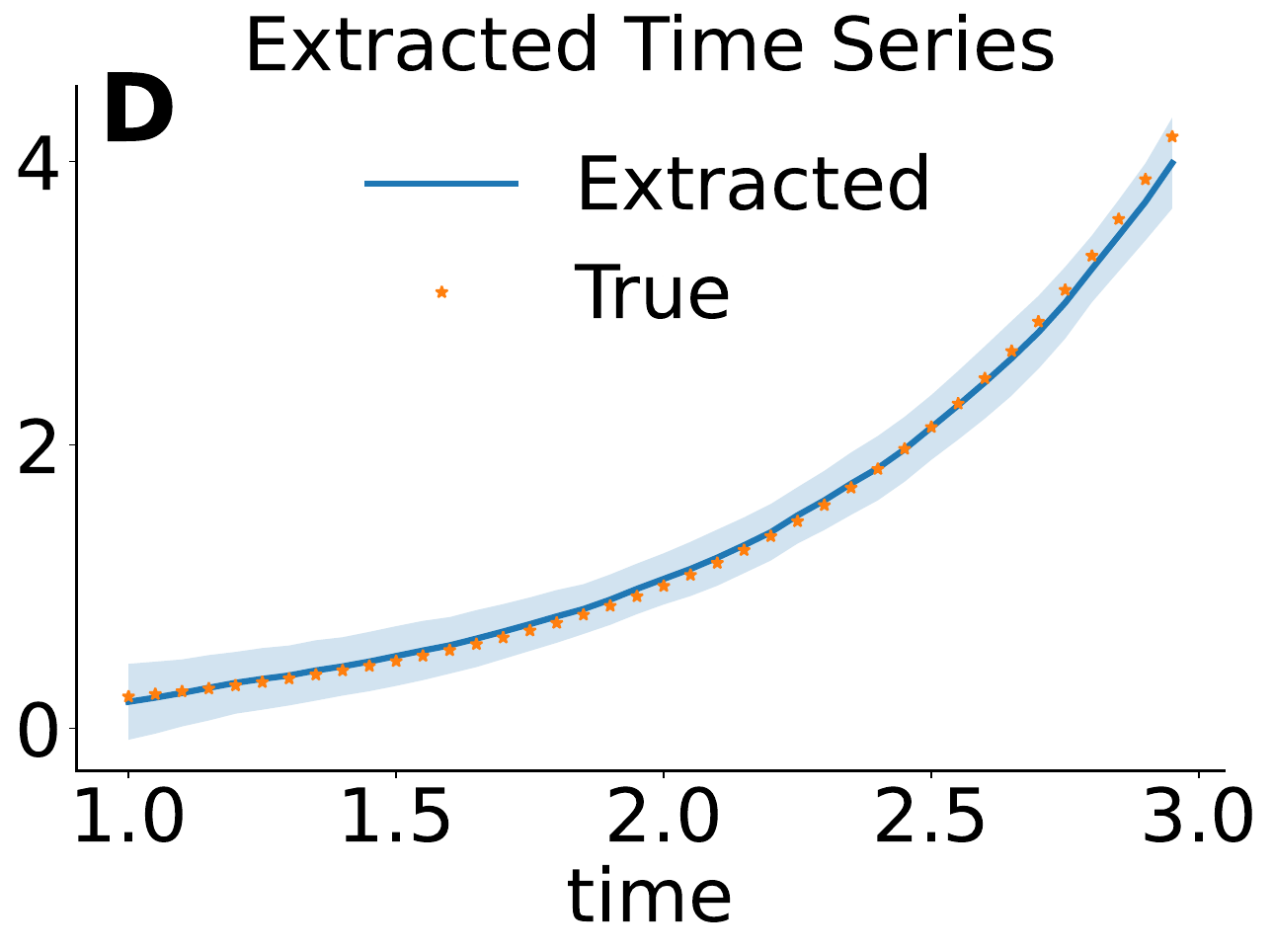}
    \caption{Problem formulation. (A) We generate a number of trajectories in the vicinity of a hyperbolic fixed point. Each trajectory is comprised of a number of growing and decaying exponentials and white noise (see panel C). (B) From these trajectories, we learn the top left eigenvector of the inferred $\A$ matrix. We use this as the time kernel of our proposed neuron. (C) We apply the time kernel to a previously unseen trajectory from the same dynamical system. (D) We extract the most dominant mode by convolving the time series with the computed time kernel. In this example, the presence of the dominant mode is clear from the projection starting at $t=1$. However, if we only look at the time series in (C), the presence of a growing mode would not be clear until around time $t=2$. Dynamics details: the time series is comprised of five different exponents with exponents given by $\{-1.5, -1, 0, 1, 1.5\}$. The time series is discretized with time-step 0.05.}
    \label{fig:sample_series}
\end{figure*}

As stated in the introduction, we assume that we are in the vicinity of a hyperbolic fixed point of the dynamics of the lag vector. This means that the dynamics of the linearized system can be decomposed as the sum of $k$ exponentials with real exponents determined by the Jacobian of $f$ at the fixed point and coefficients determined by the initial conditions of the system. We also assume that an uncorrelated observation noise (see Fig.~\ref{fig:sample_series}):
\begin{equation}\label{eq:series_form}
    x(t) = \sum_i^k c_i e^{a_i t} +\eta(t).
\end{equation}

In general, NMD extracts the exponents via an eigendecomposition, i.e. the eigenvalues of $\A$, $\lambda_i=e^{a_i}$. Since a neuron has a single output it can extract only one constituent. Because the greatest contribution to the future is given by the largest exponent (Figure~\ref{fig:sample_series}) we propose that a neuron learns the top left eigenvector of $\A$ and projects its lag vector, $\x_t$, onto this eigenvector which can then be used for prediction and control.

\subsection{Dependence of the filter on the noise level}
In this section, we look at the effect of noise on the shape of the time kernel. We first discuss the effect of additive Gaussian noise analytically, we then verify this numerically on synthetically generated data. Previous work analyzing the application of NMD to data with observation noise have primarily focused on extracting the true eigenvalues and eigenvectors of the system, that is the dynamic modes of the dynamical system in the absence of noise~\cite{noisy_dmd,stable_noisy_dmd}. Here, we are interested in extracting the most dominant mode from noisy data, and as we will see (Fig.~\ref{fig:comp2david}), the eigenvector of the noiseless system is not a good candidate for this task.

\paragraph{Analytical calculation.} Let us assume that the time series given in Eq.~\eqref{eq:series_form} has additive white observation noise, that is we assume that $\X$ satisfies Eq.~\eqref{dyn} but we observe $\X_\e=\X + \eta$ with $\eta\sim \mathcal N(0,\epsilon^2)$. We denote the noiseless data and ground truth dynamics by $\X$ and $\A$, and call the data and the inferred dynamics in the presence of noise $\X_\e$ and $\A_\e$. When averaging over the different draws of the noise, we have $\E_\eta(\X_\e\X_\e^\top) = \X\X^\top + \epsilon^2 \I_k$ and $\E_\eta(\X_{\e+}\X_\e^\top) = {\X}_+\X^\top + \epsilon^2 \S$, where $\I_k$ is the identity matrix and $\S$ is the matrix of off-diagonal ones:
\begin{equation*}
    \S=
    \begin{bmatrix}
        0 & 1 &  &   \\
         & 0 & 1 &   \\
         &  & \ddots & \ddots &   \\
         &  &  &  0 & 1 \\
        0 & 0 & \hdots & &0
    \end{bmatrix}
\end{equation*}
Note that here $\E_\eta$ denotes averaging only over the noise draws and the averaging over the samples is performed in the product of data matrices ${\X}_+\X^\top$ and ${\X}_+\X^\top$. The putative $\A_\e$ matrix (that is the $\A$ matrix that minimizes the MSE objective with noisy data $\X_\e$) is now given by
\begin{multline}
    \label{eq:A-noisy}
    \A_\e = (\X_{\e+}\X_\e^\top)(\X_\e\X_\e^\top)^{-1}\\=(\A\X\X^\top+\epsilon^2\S)(\X\X^\top+\epsilon^2\I_k)^{-1}.
\end{multline}
That is, given the ground truth dynamics and noiseless data covariances (i.e. $\A$ and $\X\X^\top$), we can determine the inferred dynamics and eigenvectors in the presence of white noise. We provide further details of this computation in the supplementary materials Sec.~\ref{app:analytic}. While the eigenvalue equations are derived analytically, they are not solvable in closed form and we look at their numerical solutions for varying noise and exponents.
Here, we summarize the results and provide some intuition.

\begin{figure*}[ht]
    \centering
    \includegraphics[height=4.95cm, trim=0 0 0 0, clip]{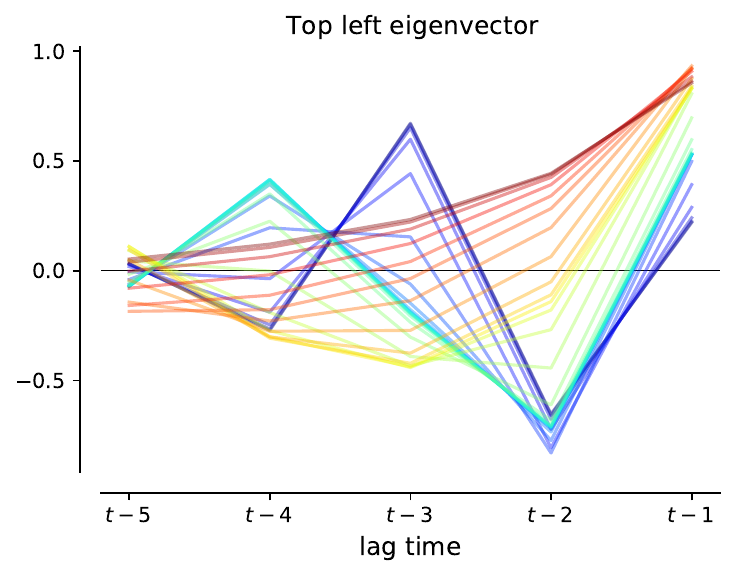}
    \hfill
    \includegraphics[height=4.95cm, trim=0 0 0 0, clip]{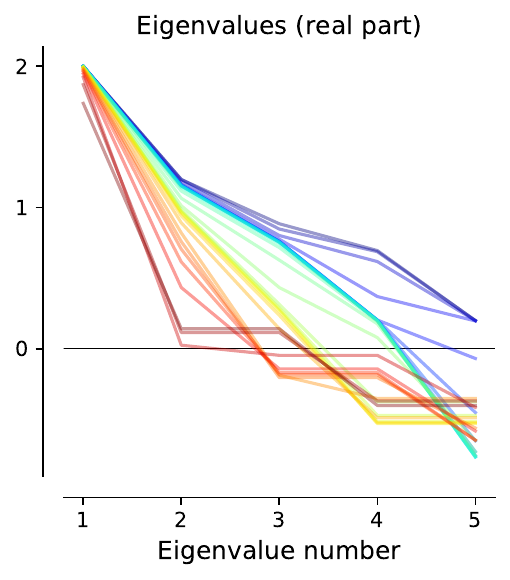}
    \hfill
    \includegraphics[height=4.95cm, trim=0 0 0 0, clip]{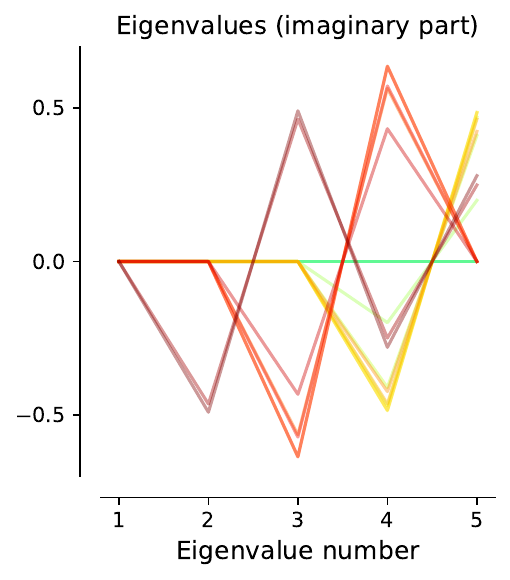}
    \hfill
    \includegraphics[height=4.9cm, trim=0 -26 0 2, ]{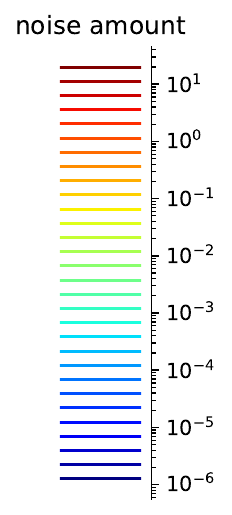}
    \caption{Analytically derived eigenvectors and eigenvalues for the lag $n=5$ for different noise levels. The $x$ axis for the eigenvalue plots denotes the eigenvalues sorted from largest to smallest real part. System with exponents $\{0.7, 0.2, -0.1, -0.4, -1.6\}$ and time-step equal to 1 for simplicity.}
    \label{fig:analytic figures0}
\end{figure*}

When looking at the eigenvalues of $\A_\e$ (Figure~\ref{fig:analytic figures0}), we see that as the noise level $\epsilon$ increases, the real parts slowly decrease and eventually cross zero. At the same time, the imaginary parts of the eigenvalues start at zero and grow when the real part becomes negative. This is intuitively understandable: when the increased amount of noise reduces the ability of the system to differentiate different modes, it starts replacing these with complex oscillatory modes.

Furthermore, at the noise value for which a mode's eigenvalue's real part crosses zero, we see that the top left eigenvector's number of phases decreases by one. This is due to the fact that the multi-phasic structure of the left eigenvectors are responsible for the bi-orthogonal structure of the left/right eigenvectors. 

\paragraph{Numerical verification.} Above we looked at the eigenvectors of $\A_\epsilon$, i.e. after averaging over different noise draws. Here, we look at these for specific noise draws, that is for specific simulations of linear dynamical systems. We looked at the response of the algorithm to synthetic data constructed from dynamics in the vicinity of a hyperbolic fixed point. We initiated a number of trajectories with different initial conditions and noise draws. Our algorithm finds the top eigenvector on the dynamics inferred from these trajectories. We then applied the learned eigenvector to a previously unseen trajectory. Figure ~\ref{fig:sample_series}(Right) shows the projection of this time series onto the largest left eigenvector and we see that we correctly recover the dominant exponential. For details of this experiment and further results and figures see supplementary materials Sec.~\ref{app:analytic}.

\begin{figure*}[ht]
     \centering
     \begin{subfigure}[b]{0.32\textwidth}
         \centering
         \includegraphics[width=\textwidth, trim=0 1 13 10, clip]{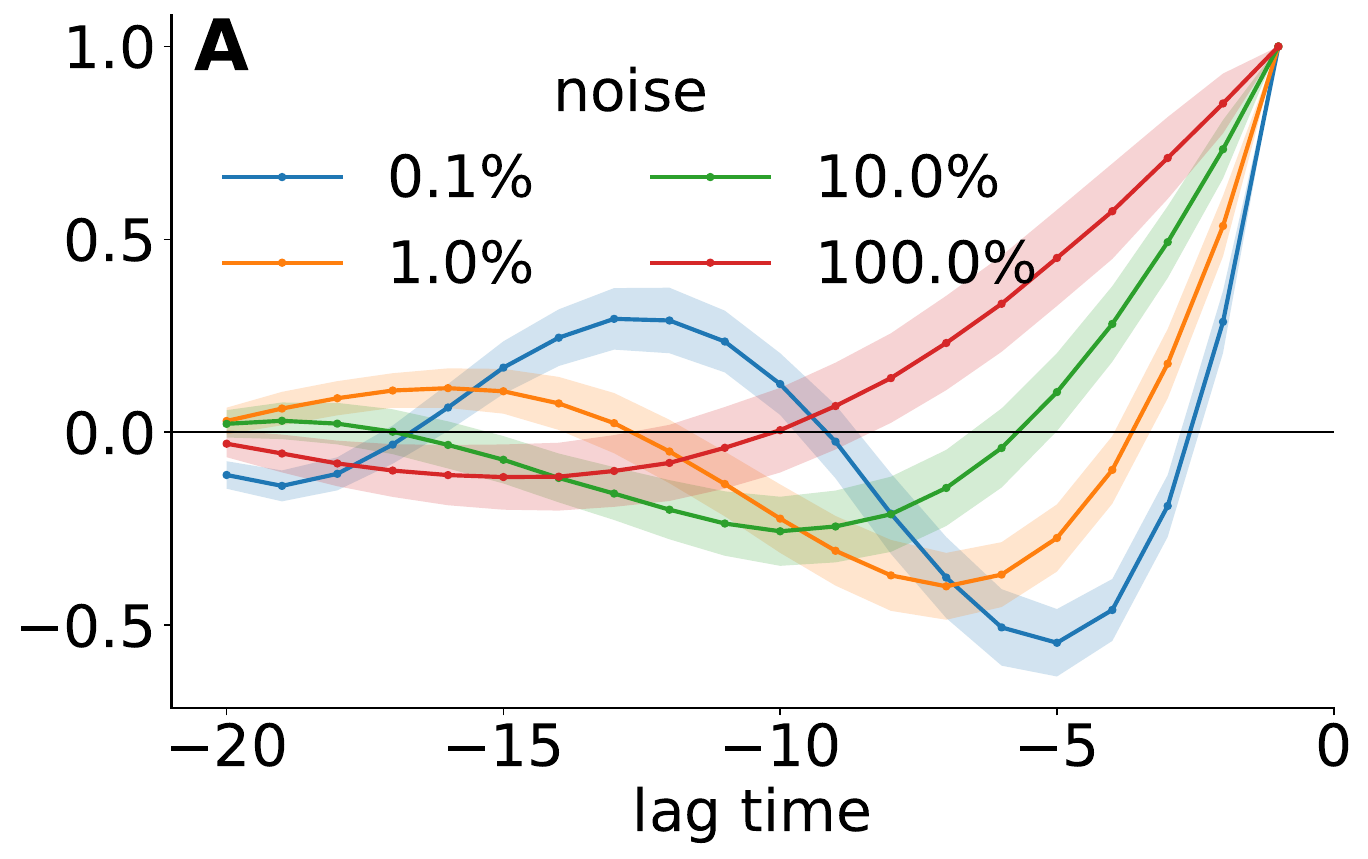}
         \label{fig:synth_vs_noise}
     \end{subfigure}
     \hfill
     \begin{subfigure}[b]{0.32\textwidth}
         \centering
         \includegraphics[width=\textwidth, trim=0 1 13 10, clip]{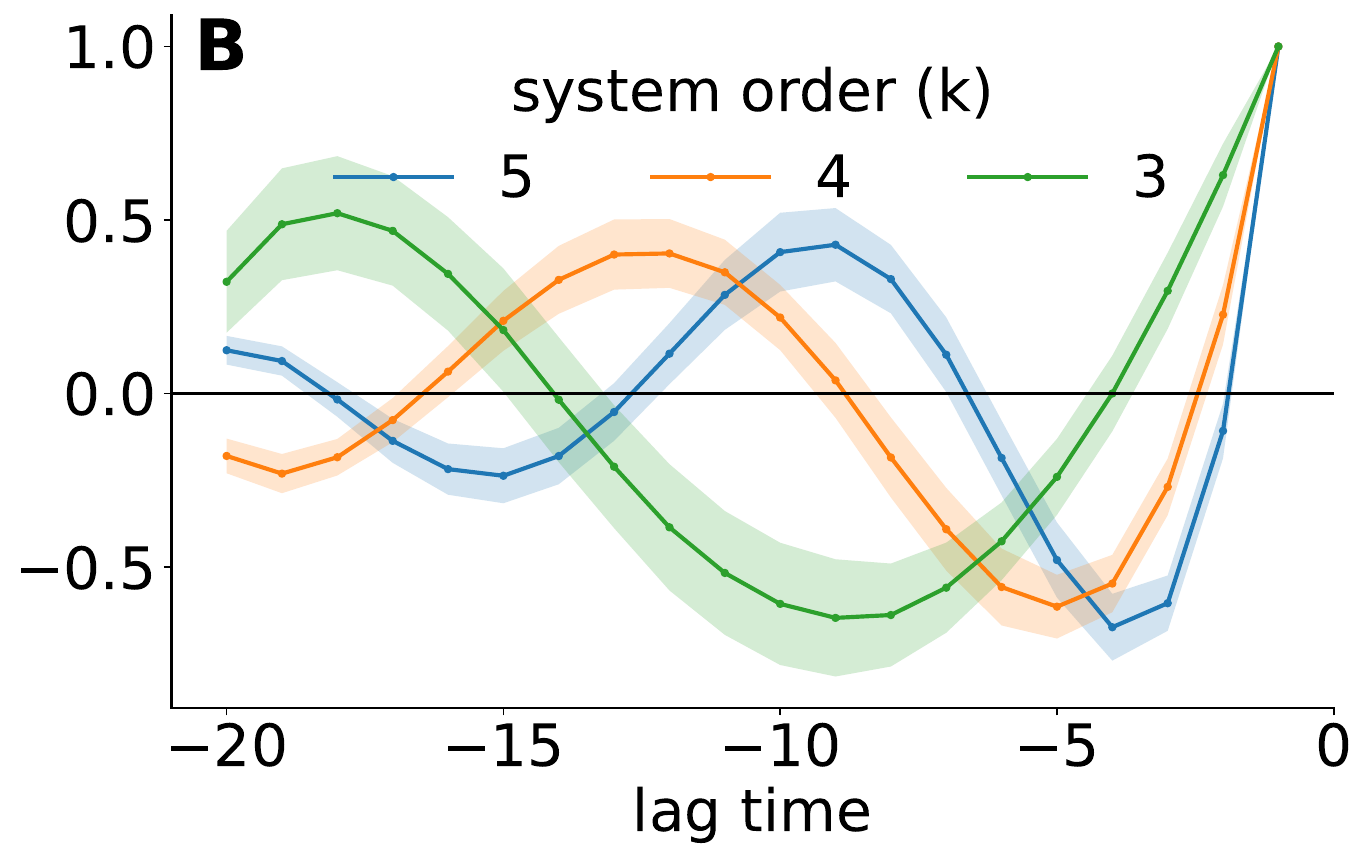}
         \label{fig:synth_vs_order}
     \end{subfigure}
     \hfill
     \begin{subfigure}[b]{0.32\textwidth}
         \centering
         \includegraphics[width=\textwidth, trim=0 1 13 10, clip]{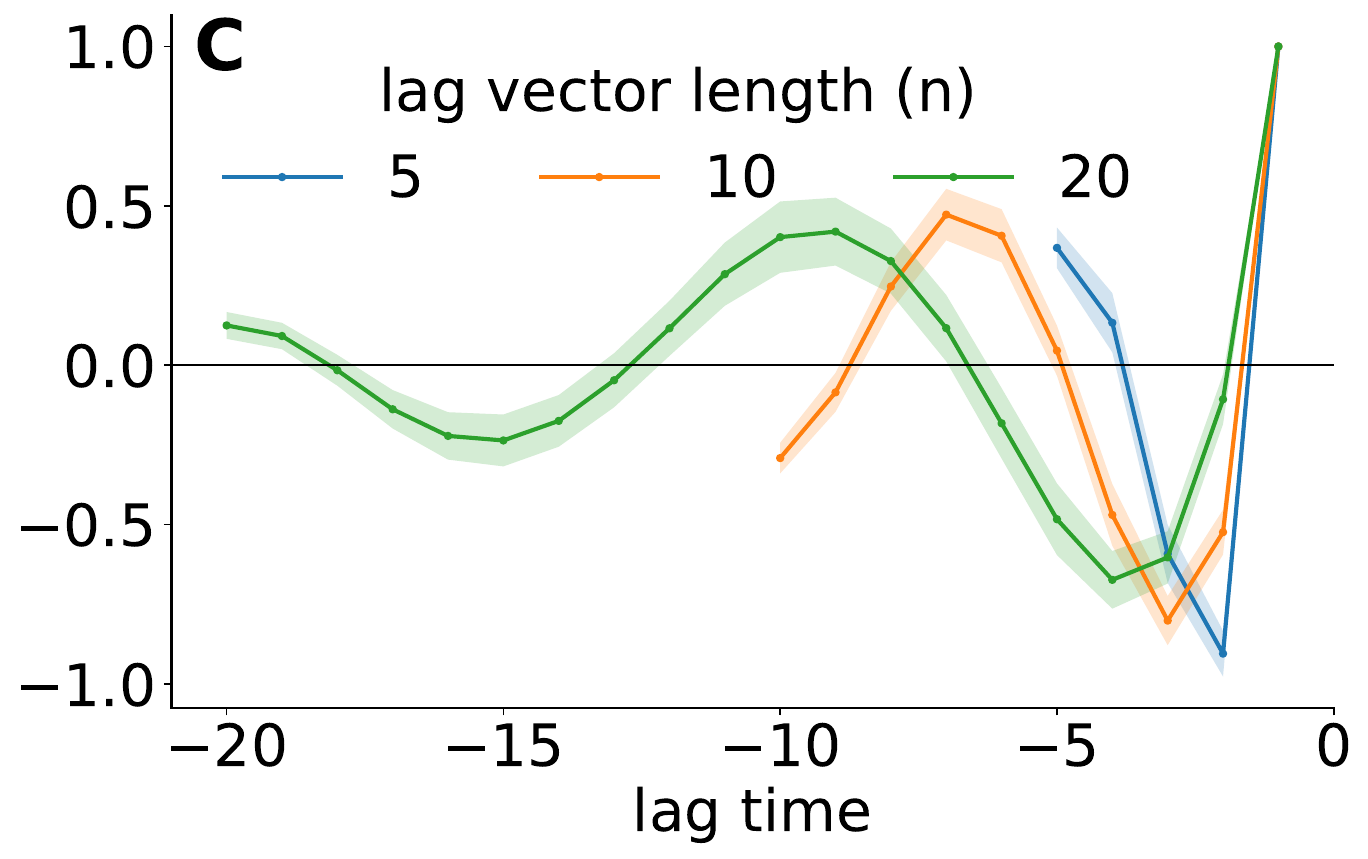}
         \label{fig:synth_vs_length}
     \end{subfigure}
        \caption{Dependence of the filter shape for the rank 5 system in Fig.~\ref{fig:sample_series} on system parameters: (A) noise amplitude, (B) system order (the number of exponential constituents, $k$), and (C) length of the lag vector, $n$. In (B) and (C) the noise is fixed at $0.01\%$.  In (B), to achieve a comparable system of lower rank $k$, we keep the top $k$ exponents of the system.}
        \label{fig:synth_results}
\end{figure*}

Figure~\ref{fig:synth_results} shows the effect of various parameters such as the noise, the system order, and the lag vector length on the shape of the filter. We see that adding noise decreases the number of the phases as expected (Fig.~\ref{fig:synth_results}A). Similarly as we decrease the system order $k$, the number of phases decrease  (Fig.~\ref{fig:synth_results}B) as expected from the bi-orthogonality argument given at the end of Sec.~\ref{sec:DMD}.  Figure~\ref{fig:synth_results}C shows that with a finite amount of noise, as the length of the lag vector is increased, the number of phases can slowly increase. This is due to the fact that with longer lag vectors, there is higher noise tolerance. Indeed as we take the noise values to zero, we see that the lag vector length no longer affects the number of phases in the filter. Increasing the length of the lag vector merely stretches the filter shape (Fig.~\ref{fig:vslength_lownoise} in the supplementary materials).

\subsection{Comparison with optimal projection under noisy observations}

To verify that in the presence of noise our algorithm provides a sensible approximation of the dominant dynamical mode, in this section we discuss a noise-optimal projection alternative and provide empirical comparisons.

\begin{figure}
    \centering
    \includegraphics[width=0.45\textwidth]{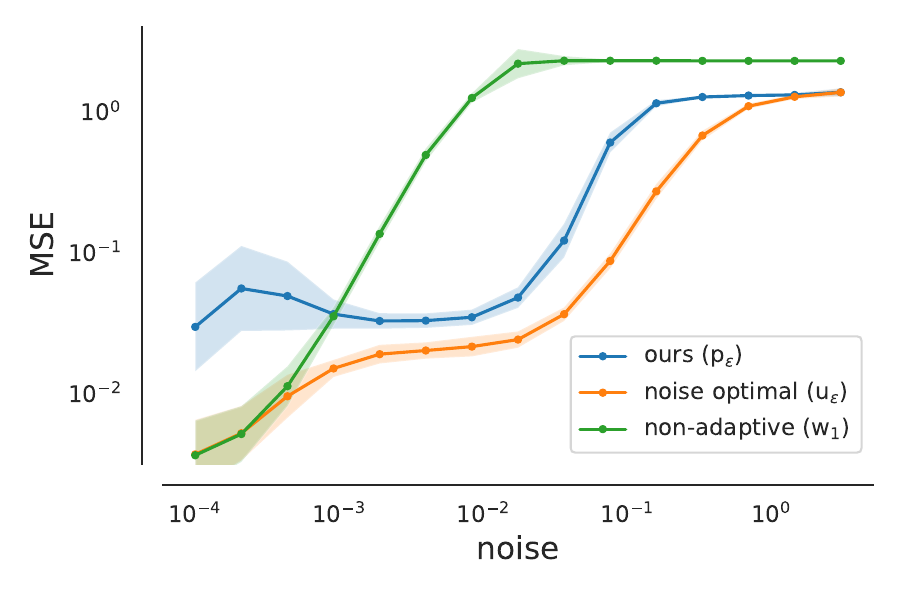}
    \caption{Comparison of the performance with respect to the mean square error between the recovered dominant exponential and the ground truth. The `noise optimal' solution is given by the filter in Eq.~\eqref{eq:noise_optimal} and `noiseless filter' refers to using the filter compted in the absence of noise but applied to noisy observations. The system is the same as in Fig.~\ref{fig:sample_series}. }
    \label{fig:comp2david}
\end{figure}

Let $\x_t$ be a time series and suppose our goal is to extract the dominant exponential mode by projecting $\x_t$ onto the filter given by $\v$ (derived as the top eigenvector of \eqref{eq:gen_eig_prob} when there is no observation noise). We refer to $\v$ as the \emph{noiseless filter}. Suppose, however, we do not have access to $\x_t$, but only a noisy observation $\y_t=\x_t+\n_t$, where $\n_t$ is isotropic Gaussian noise with variance $\alpha^2$. What is the optimal projection of $\y_t$? Consider the optimization problem
\begin{multline*}
    \min_{\w} E[|\w^\top\y_t-\v^\top\x_t|^2]=\\
    \min_\w (\w-\v)^\top\C_{xx}(\w-\v)+\alpha^2\w^\top\w,
\end{multline*}
where we have used the fact that $\n_t$ is mean zero isotropic noise independent of $\x_t$. Solving for the optimal $\w$ we find:
\begin{align}
    \w=(\C_{xx}+\alpha^2\I)^{-1}\C_{xx}\v.
    \label{eq:noise_optimal}
\end{align}
In order to apply this method, we need to know the noise variance $\alpha$ and also the noise-free projector $\v$ and covariance $\C_{xx}$. Computing these noise-free quantities is challenging, especially in a problem where the dynamics might be changing over time. Furthermore, computing the full covariance matrix $\C_{xx}$ requires more samples than just computing the top subspace in our generalized eigenvalue problem (Eq.~\eqref{eq:gen_eig_prob}). In practice, we find that even if we know $\alpha$ and $\v$, our method gives comparable performance to this noise-optimal solution (Fig.~\ref{fig:comp2david}). Furthermore, our method provides a better estimate of the most dominant mode than the non-adaptive method which uses the true (i.e. noiseless) eigenvector. In practice, in order to compute the noiseless filter $\v$, we set the noise variance $\alpha=10^{-6}$.

\section{Discussion}
We propose to model a neuron on an algorithmic level as performing a NMD on its input. Specifically, the temporal filter of a neuron is given by the top left eigenvector of the generalized eigenproblem formulated in terms of covariances of lag vectors. The neuron outputs the fastest growing mode present in the input thus predicting a future trend.

How could a biological neuron find the top eigenvector of the generative matrix? One possibility is for a neuron to solve the generalized eigenproblem \eqref{eq:gen_eig_prob} using a power iteration. Specifically, this would require two types of history-dependent active conductances (ion channels) which encode input covariances, $\mathbf{X_+X^\top}$ and $\mathbf{XX^\top}$, with opposite signs (depolarizing and hyperpolarizing). Then the solution of \eqref{eq:gen_eig_prob} would be given by the voltage that balances the two currents. The output of a neuron would reflect such voltage, thus projecting the input on the top left eigenvector of the transition matrix.

Whereas a single neuron computing a scalar signal can represent only one eigenmode, multiple neurons may extract multiple modes (not just the top eigenmode). For each neuron to represent a different mode their output must be decorrelated, for example, by the lateral inhibitory connections. The strengths of such connections can be adjusted using biologically plausible local learning rules as has been shown by some of the authors previously in the context of extracting eigenmodes of the multichannel covariance \cite{pehlevan2015hebbian,pehlevan2017similarity,pehlevan2019neuroscience}. Also, we proposed a framework for deriving multichannel neural networks for solving symmetric generalized eigenvalue problems \cite{PRXLife.1.013008} and the approach can potentially be extended to solve non-symmetric generalized eigenvalue problems. We anticipate that a similar approach can be applied to extracting eigenmodes of the time-lagged dynamics using local learning rules.

Because the matrix $\A$ in Eq.\eqref{dyn} is generically non-normal, its eigenvectors are not guaranteed to be orthogonal. In practice, this means that they are not robust to perturbations to the elements of the matrix $\A$. This requires particular care to ensure that the eigenvalues are sufficiently distinct as determined by the pseudospectrum of $\A$ \cite{TrefethenEmbree+2020}.  

Neuronal projection of its input onto the subspace corresponding to the fastest growing mode has an interpretation in terms of the phase portrait of the generative dynamical system. Close to a hyperbolic fixed point the dynamics are linear and the fastest growing mode would correspond to the unstable invariant subspace which evolves into an attracting manifold away from the fixed point. Therefore, the sign of the projection onto the unstable subspace predicts along which unstable manifold the future trajectory of the system will develop beyond the linear regime. If the neuronal output is rectified, it effectively assigns the trajectory to a particular future state. Output of a layer of such rectifying neurons becomes a latent vector variable that can in turn be an input to the next layer. This opens a path to stacking the layers of such neurons to discover more and more abstract latent variables. 



\bibliography{references.bib}
\bibliographystyle{unsrt}




\clearpage
\appendix

\begin{center}\LARGE{\textbf{Supplementary Materials}}\end{center}
\vspace{3pt}
This is the supplementary materials section for the Physical Review Research submission titled ``Neuronal Temporal Filters as
Normal Mode Extractors''.

\section{Details of analytic arguments}\label{app:analytic}

Starting with the equation~\eqref{eq:A-noisy} describing the effect of noise:
\begin{equation}
    \A = (\X_+\X^\top)(\X\X^\top)^{-1}=(\A_0\X_0\X_0^\top+\epsilon^2\S)(\X_0\X_0^\top+\epsilon^2\I_k)^{-1}.
\end{equation}
we can derive the noisy version of the dynamics if we know the covariance structure of the non-noisy data $\X_0\X_0^\top$. It is intuitively understandable why this structure is necessary since the effect of noise depends not just on the eigenvalues and eigenvectors of $\A$, it also depends on the magnitudes of each individual mode. Specifically, if we assume the generative model in Eq.~\ref{eq:series_form}, it is not just the $a_i$ that determine the noisy dynamics, the $c_i$ are also important and this information is not present in $\A$ but is available in  $\X_0\X_0^\top$.

For plots given in Fig.~\ref{fig:analytic figures0}, we computed the $\X_0\X_0^\top$ using time series of the form in Eq.~\ref{eq:series_form}. We take the length of the time series, to be just long enough to compute a full rank covariance matrix.


\section{Other Figures}\label{app:figs}
\begin{figure}[t]
    \centering
    \includegraphics[width=0.3\textwidth]{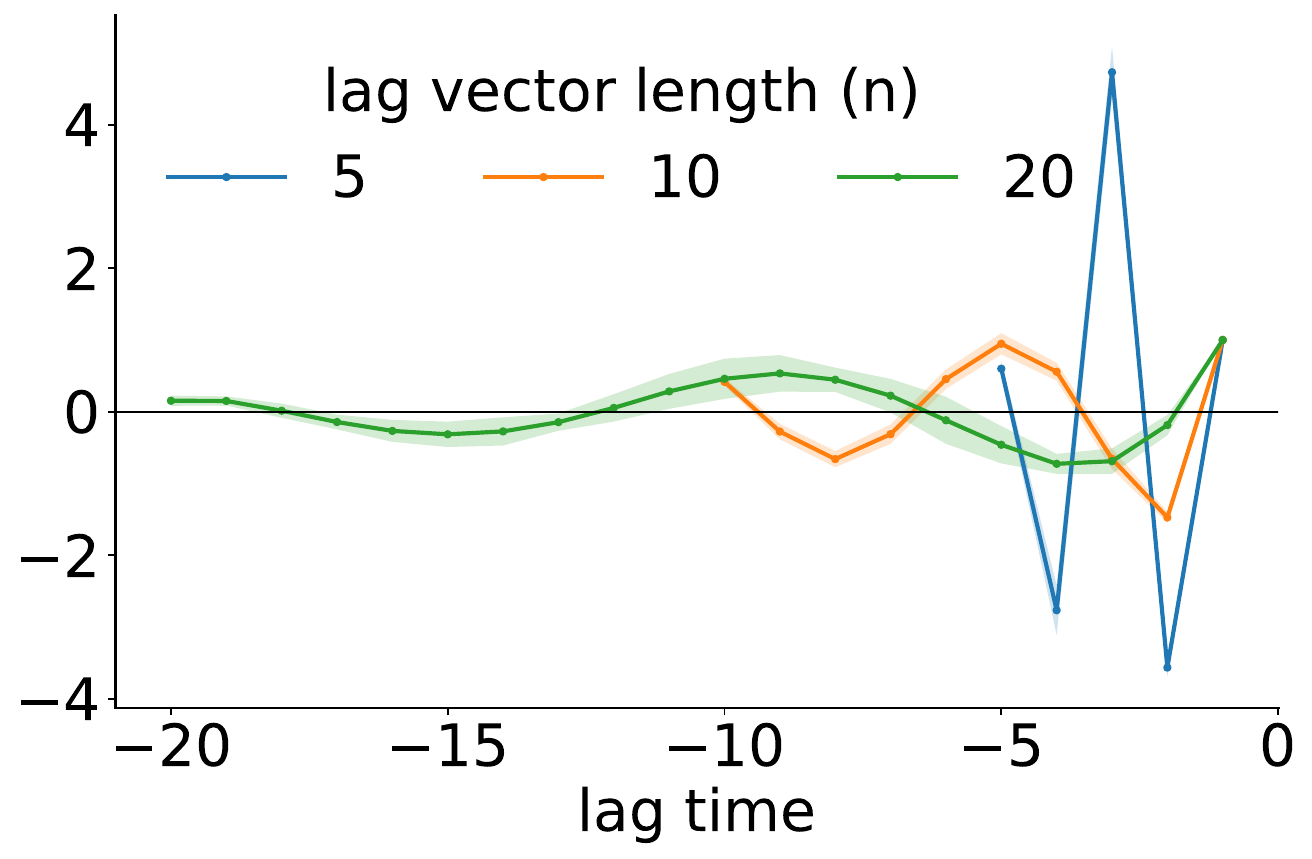}
    \caption{The dependence of the filter shape vs the length of the lag vector at low noise values ($0.00001\%$).}
    \label{fig:vslength_lownoise}
\end{figure}

Figure~\ref{fig:vslength_lownoise} shows the dependence of the filter shape as we vary the lag vector length for a small amount of noise. We see that increasing the lag vector length no longer affects the number of phases (cf. Fig.~\ref{fig:synth_results}C).

\section{Inverse of a Vandermonde matrix}\label{app:vandermonde}

Recall that in the noiseless case (i.e., $\epsilon=0$), the matrix $\A$ has eigendecomposition $\A=\V\Lam\V^{-1}$, where $\V$ is a Vandermonde matrix. In this section we show that the top left eigenvector of a Vandermonde matrix of rank $n$ changes sign $n-1$ times. Consider the Vandermonde matrix
\begin{align*}
    \V:=\begin{bmatrix}
        1 & 1 & \cdots & 1 \\
        \lambda_1 & \lambda_2 & \cdots &\lambda_n\\
        \vdots & \vdots & & \vdots\\
        \lambda_1^{n-1}&\lambda_2^{n-1}&\cdots&\lambda_n^{n-1}
    \end{bmatrix}
\end{align*}
where $\lambda_1>\cdots>\lambda_n>0$. Rawashdeh \citep{rawashdeh2019simple} showed that the inverse of the Vandermonde matrix $V$ is given by
\begin{align*}
    (\V^{-1})_{ij}=(-1)^{i+j}S_{n-j,i}\frac{\prod_{k<l\text{ s.t.\ }l,k\ne j}^n(\lambda_k-\lambda_l)}{\prod_{k<l}^n(\lambda_k-\lambda_l)},
\end{align*}
where
\begin{align*}
    S_{j,k}:=\sum_{1\le i_1<\cdots<i_k\le n,i_\ell\ne j}\prod_{m=1}^k\lambda_{i_m}>0.
\end{align*}
Importantly, the sign of the $(1,j)$\textsuperscript{th} entry is positive (resp.\ negative) if $j$ is odd (resp.\ even). Therefore, the top left eigenvector changes sign $n-1$ times.

\section{Imaginary Parts in Eigenvalues}
\begin{figure*}[t]
    \centering
    \includegraphics[width=0.24\textwidth, trim=5 0 5 0]{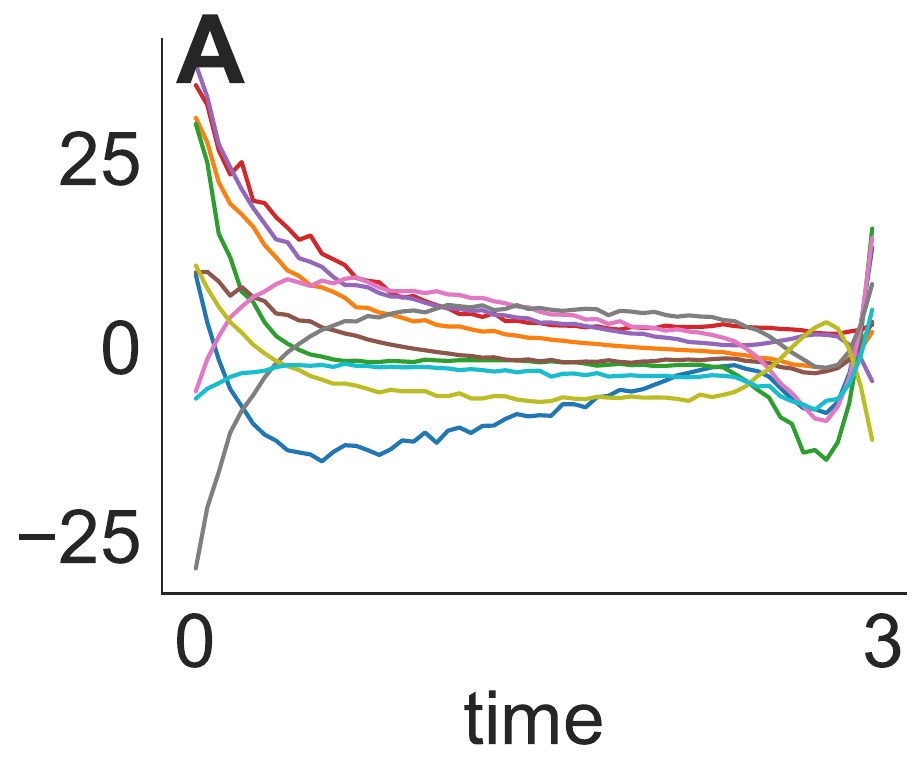}
    \hfill
    \includegraphics[width=0.23\textwidth, trim=5 0 5 0]{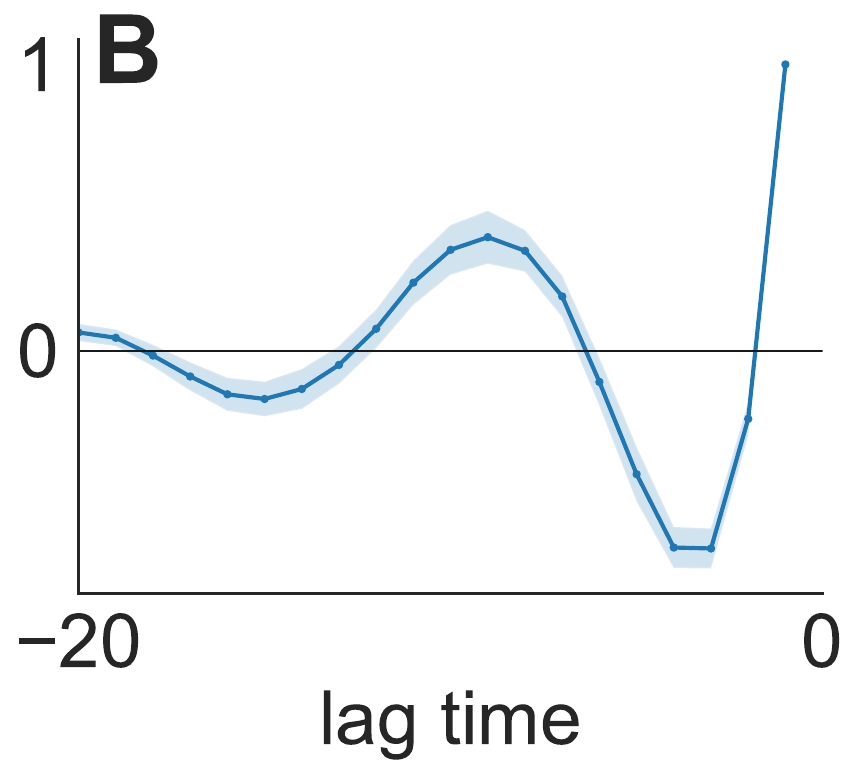}
    \hfill
    \includegraphics[width=0.24\textwidth, trim=5 0 5 0]{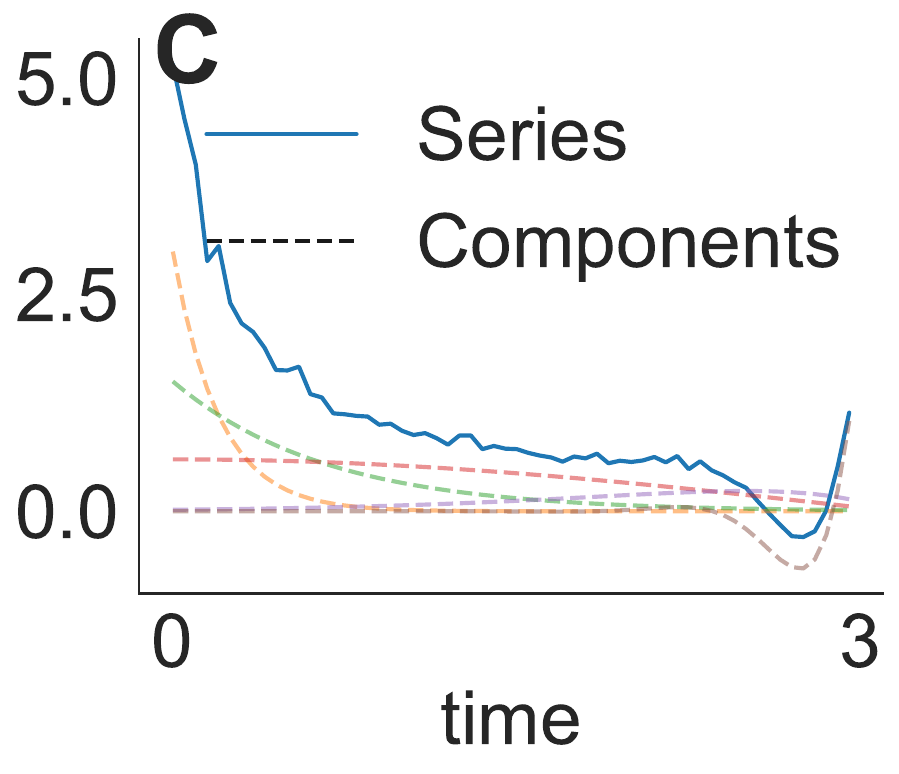}
    \hfill
    \includegraphics[width=0.24\textwidth, trim=5 0 5 0]{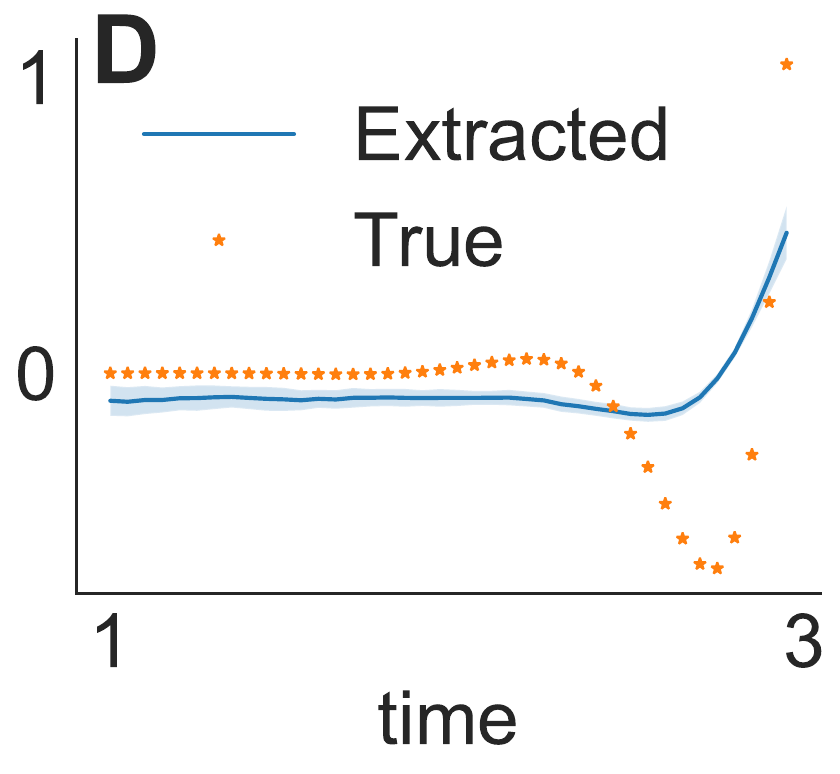}
    \\
    \includegraphics[width=0.24\textwidth, trim=5 0 5 0]{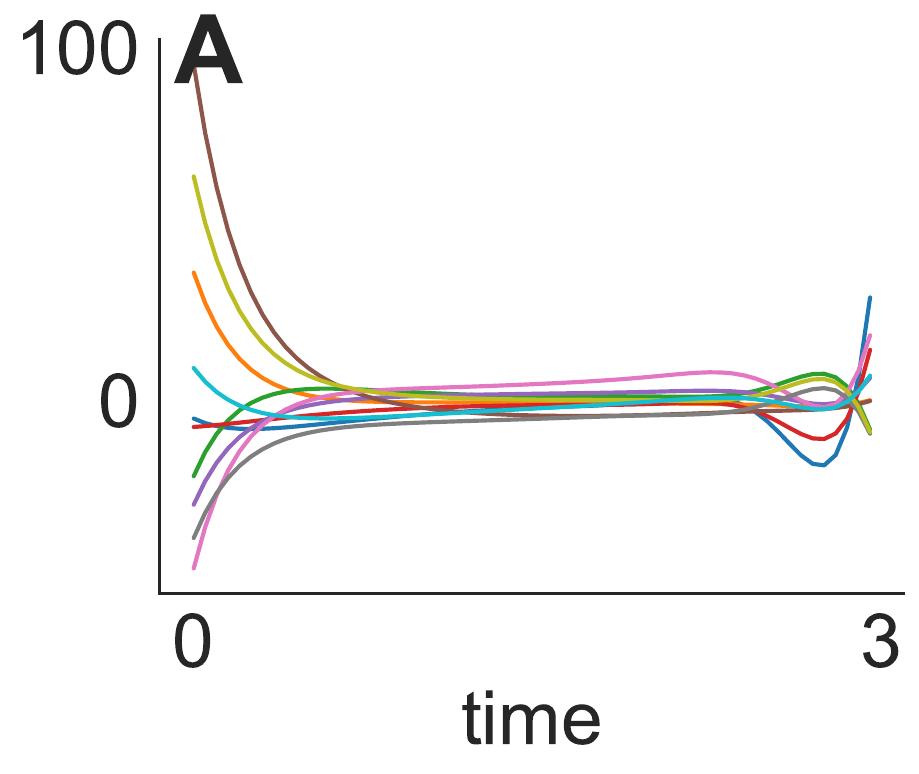}
    \hfill
    \includegraphics[width=0.23\textwidth, trim=5 0 5 0]{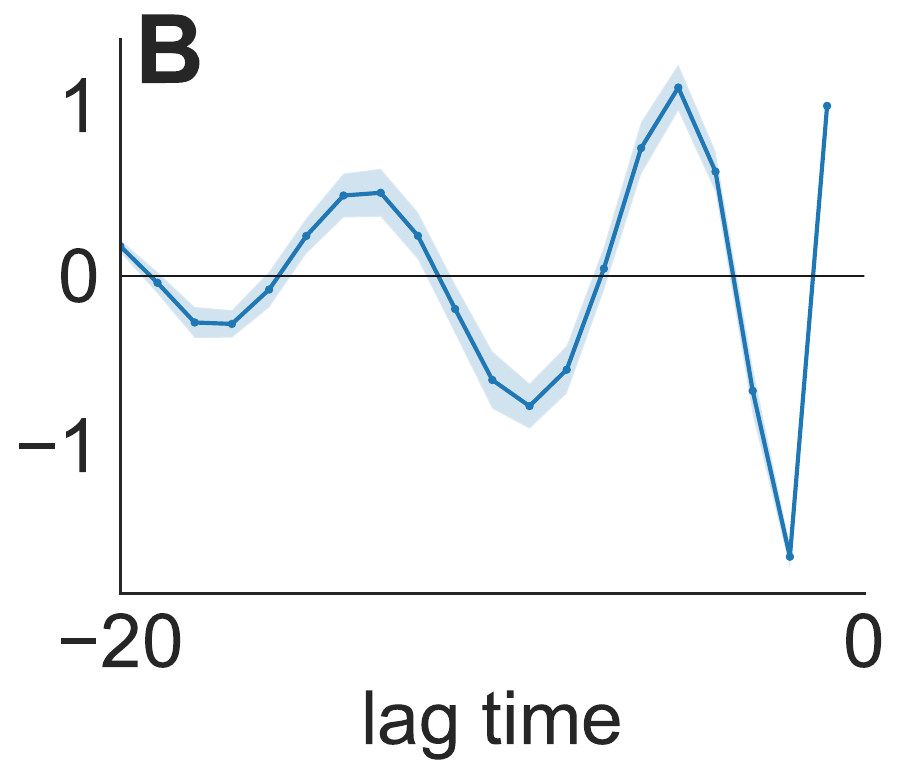}
    \hfill
    \includegraphics[width=0.24\textwidth, trim=5 0 5 0]{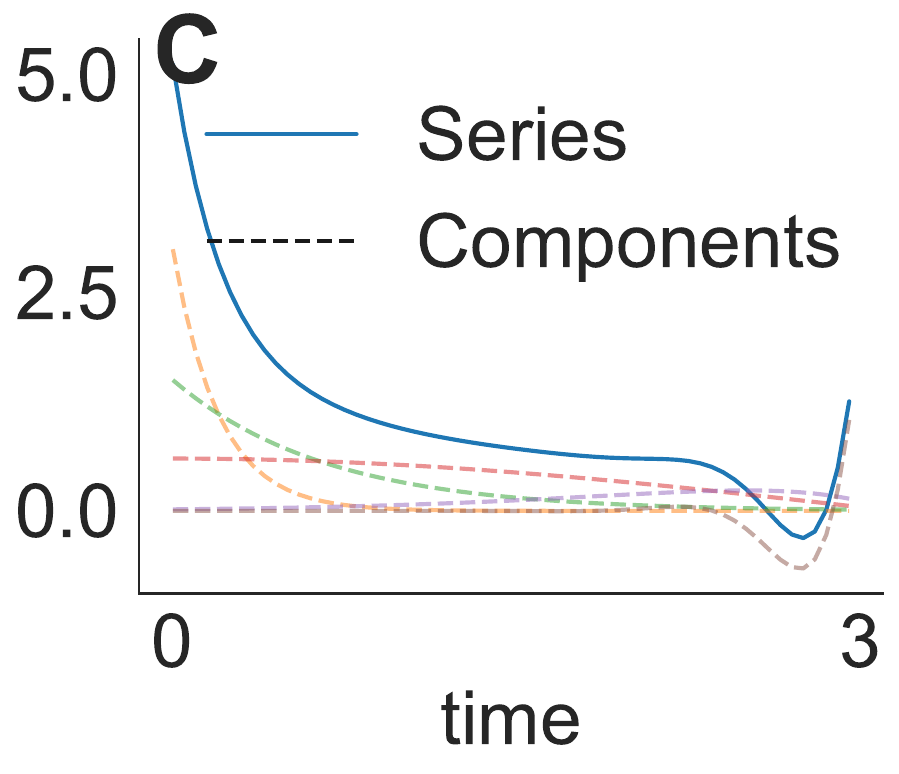}
    \hfill
    \includegraphics[width=0.24\textwidth, trim=5 0 5 0]{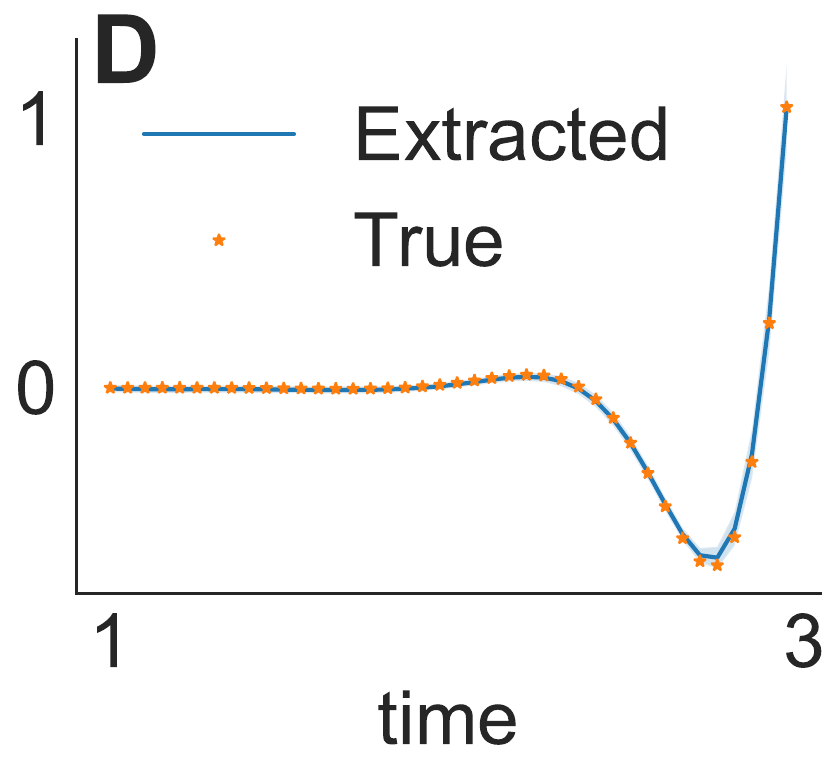}
    \\
    \includegraphics[width=0.24\textwidth, trim=5 0 5 0]{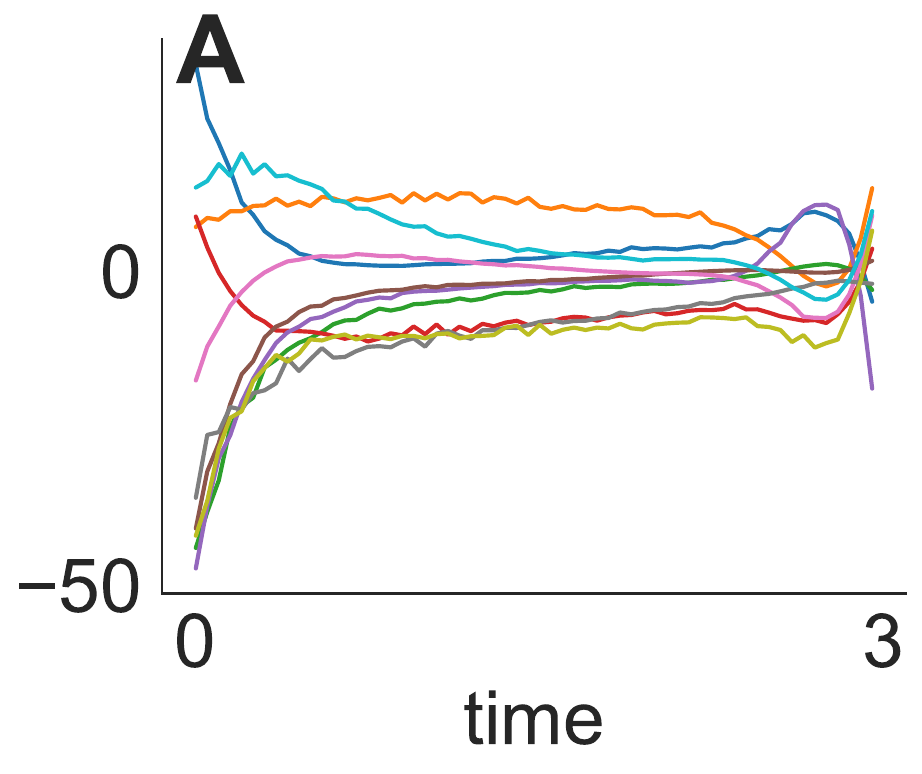}
    \hfill
    \includegraphics[width=0.23\textwidth, trim=5 0 5 0]{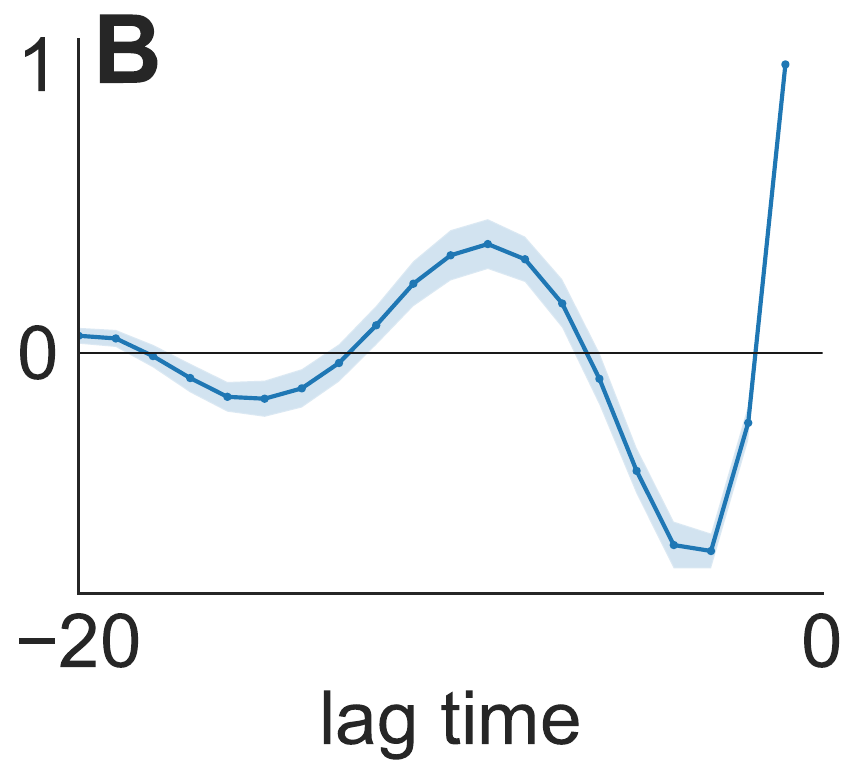}
    \hfill
    \includegraphics[width=0.24\textwidth, trim=5 0 5 0]{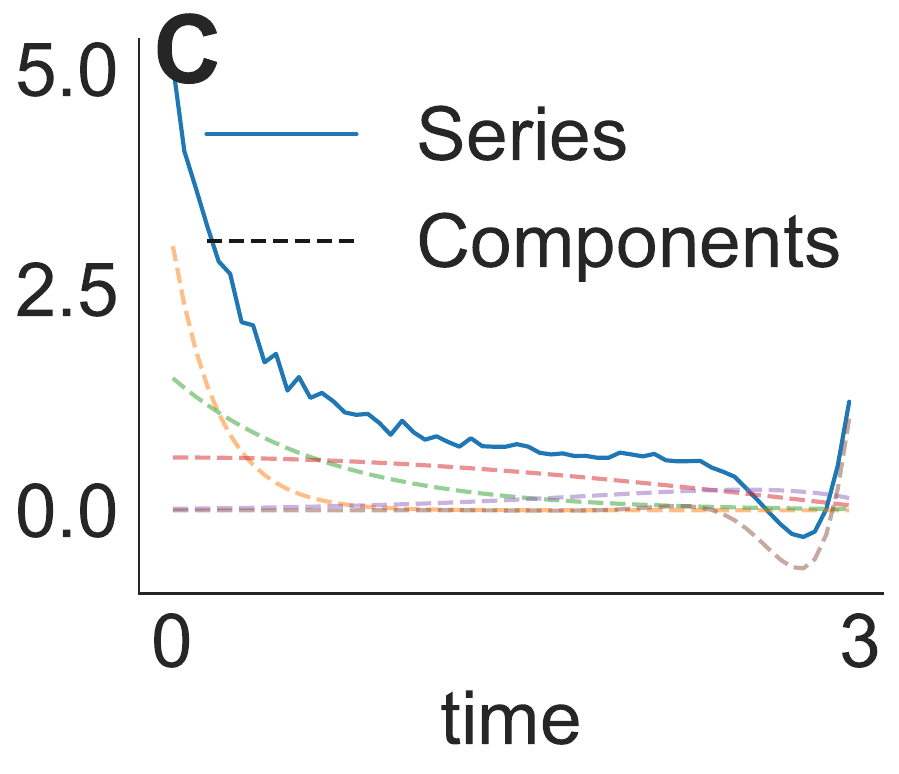}
    \hfill
    \includegraphics[width=0.24\textwidth, trim=5 0 5 0]{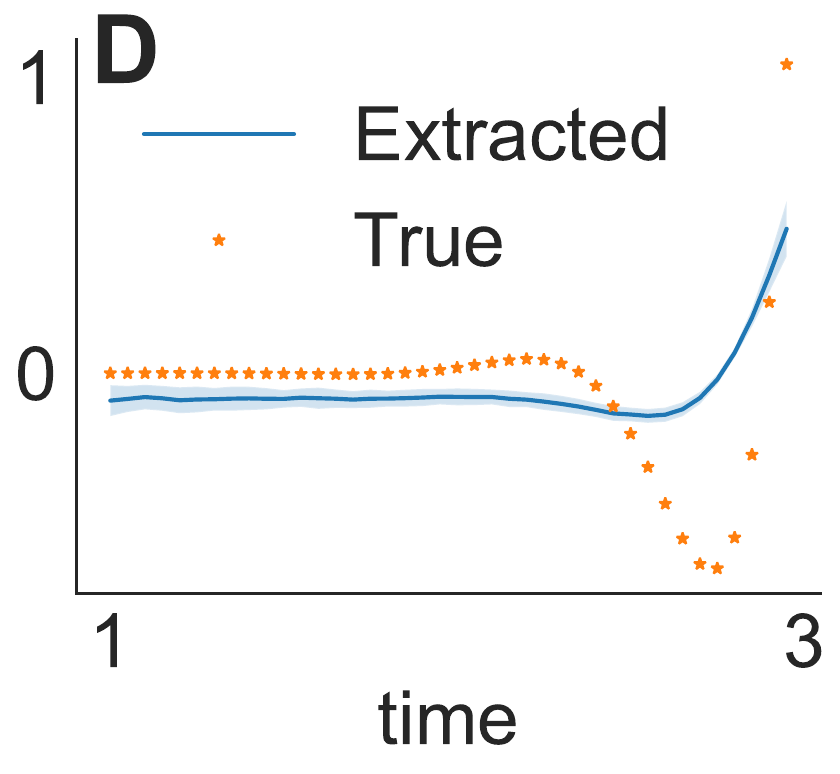}

        \caption{We reproduce Figure \ref{fig:sample_series} with imaginary parts in the exponents. (A) Imaginary parts in all trajectories except the one with the largest eigenvalue. (B) Imaginary parts in all trajectories including the one with the largest eigenvalue. (C) Imaginary parts in all trajectories including the one with the largest eigenvalue, but lower noise level.}
            \label{fig:imag_nolarge_same}
\end{figure*}

In the main text we only examine a time series composed of real exponents. Here we examine the case where exponents have imaginary parts which creates oscillatory behavior in the training and test time series. We reproduce Figure \ref{fig:sample_series} where we infer the $\A$ matrix from training trajectories, extract the top left eigenvector and use it to extract the dominant mode from an unseen time series. In Figure \ref{fig:imag_nolarge_same} (Row A) we add an imaginary part in all trajectories except the one with the largest eigenvalue. We see that we are still able to accurately extract the largest mode of the time series. In Figure \ref{fig:imag_nolarge_same} (Row B) we add an imaginary part in all trajectories including the one with the largest eigenvalue. Here we fail to accurately reconstruct the largest mode, although we get the qualitative behavior correct. In Figure \ref{fig:imag_nolarge_same} (Row C) we repeat the test in Row B but in addition we lower the noise level. Under these circumstances we see that we are able to accurately reconstruct the largest mode, even when it contains an imaginary part. The specific parameters are given in Table \ref{tbl:imgtable}.

\begin{table}
  \centering
  \begin{tabular}{l|l|l}
    \toprule
    \cmidrule(r){1-2}
    Row  & Exponent Imaginary Part  & noise \\
    \midrule
    A & $\{-5+0.2i, -1.5+7i, 1i, 1.5+0.1i, 5.001\}$ &  $0.05$ \\
    B & $\{-5+0.4i, -1.5+0.2i, 0.5i, 1.5+0.5i, 5.001+6i\}$ &  $0.05$ \\
    C & $\{-5+0.4i, -1.5+0.2i, 0.5i, 1.5+0.5i, 5.001+6i\}$ &  $0.0001$ \\
    \bottomrule
  \end{tabular}
  \caption{Experimental parameters for Figure \ref{fig:imag_nolarge_same}}
\label{tbl:imgtable}
\end{table}

\clearpage

\end{document}

%% file: commands.tex
\newcommand{\n}{{\bf n}}

\newcommand{\w}{{\bf w}}
\newcommand{\x}{{\bf x}}
\newcommand{\y}{{\bf y}}

\newcommand{\A}{{\bf A}}

\newcommand{\C}{{\bf C}}

\newcommand{\E}{\mathbb{E}}
\newcommand{\F}{{\bf F}}

\newcommand{\I}{{\bf I}}

\newcommand{\R}{\mathbb{R}}
\renewcommand{\S}{{\bf S}}

\newcommand{\V}{{\bf V}}

\newcommand{\X}{{\bf X}}

\renewcommand{\v}{{\bf v}}

\newcommand{\e}{{\epsilon}}

\renewcommand{\S}{{\bf S}}

\newcommand{\Lam}{\boldsymbol{\Lambda}}